\documentclass[aps,a4paper,12pt,preprint,pre,longbibliography]{revtex4-1}
\usepackage{graphicx}
\usepackage{amsmath}

\usepackage{amssymb}

\usepackage{amsfonts}
\usepackage[usenames]{color}
\usepackage[applemac]{inputenc}
\usepackage{ulem}
\def\refpar[#1]{(\ref{#1})}
\def\blue[#1]{\textcolor{blue}{#1}}
\definecolor{vert}{RGB}{51,110,23}

\newcommand{\beq}{\begin{equation}}
\newcommand{\eeq}{\end{equation}}
\newcommand{\beqa}{\begin{eqnarray}}
\newcommand{\eeqa}{\end{eqnarray}}
\newcommand{\ba}{\begin{array}}
\newcommand{\ea}{\end{array}}
\newcommand{\dd}{\mathrm{d}}

\newcommand{\donc}{\quad \Longrightarrow \quad}
\newcommand{\K}{\widetilde{K}}
\newcommand{\lag}{{\cal L}}
\newcommand{\OG}{{\cal O}}

\newcommand{\xt}{\widetilde{x}_0}
\newcommand{\yt}{\widetilde{y}_0}
\newcommand{\dxt}{\dot{\widetilde{x}}_0}
\newcommand{\dyt}{\dot{\widetilde{y}}_0}

\begin{document}

\title{Parametric resonance in a  conservative system of coupled nonlinear oscillators.}
\author{Johann Maddi}\author{Christophe Coste}\author{Michel Saint Jean}\affiliation{ Laboratoire "Mati\`ere et Syst\`emes Complexes" (MSC), UMR 7057 CNRS, Universit\'e Paris--Diderot (Paris 7),
75205 Paris Cedex 13, France
}

\date{\today}
\begin{abstract}

We study a conservative system of two nonlinear coupled oscillators. The eigenmodes of the system are thus nonlinearly coupled, and one of them may induce a parametric amplification of the other, called an autoparametric resonance of the system. The parametric amplification implies two time scales, a fast one for the forcing and a slow one for the forced mode, thus a multiscale expansion is suitable to get amplitude equations describing the slow dynamics of the oscillators. We recall the parametric resonance in a dissipationless system, the parametrically forced Duffing oscillator, with emphasis on the energy transfer between the oscillator and the source that ensures the parametric forcing. Energy conservation is observed when averaging is done on the slow time scale relevant to parametric amplification,evidenced by a constant of the motion in the amplitude equation. Then we study a dimer in a periodic potential well, which is a conservative but non integrable system. When the dimer energy is such that it is trapped in neighboring potential wells, we derive coupled nonlinear differential equations for the eigenmodes amplitudes (center of mass motion and relative motion). We exhibit two constants of the motion, which demonstrates that the amplitude equations are integrable. We establish the conditions for autoparametric amplification of the relative motion by the center of mass motion, and describe the phase portraits of the system. In the opposite limit, when the dimer slides along the external potential so that the center of mass motion is basically a translation, we calculate the amplitude equation for the relative motion. In this latter case, we also exhibit autoparametric amplification of the relative motions of the dimer particles. In both cases, the comparison between numerical integration of the actual system and the asymptotic analysis evidences an excellent agreement. 

PhySH~: Discipline~: Nonlinear Dynamics; Research areas : Classical Mechanics; Physical systems~: Coupled Oscillators; Techniques~: Phase space methods.
\end{abstract}

\maketitle

\section{Introduction}

The dynamics of coupled linear oscillators is easily described as a superposition of their eigenmodes motions. This is not the case for nonlinear oscillators, since their characteristic frequencies depend on the amplitude of the oscillations. The coupling may therefore induce a resonant response of some of the oscillatory modes. Nonlinear resonances are very common in physical systems, and usually difficult to analyse~\cite{Kartashova10}. 

In a conservative system, without any external source, there could be a resonant energy exchange between two oscillatory modes. Recent studies have considered coupled Duffing oscillators~\cite{Denardo99,Naz11,Sabarathinam13,Lenci22} that exhibit nonlinear resonant response. A swinging spring is another well known mechanical apparatus that may evidence resonant behavior when the frequencies of the elastic and pendular oscillations are in the ratio 2:1~\cite{Kuznetsov99,Lynch04}.  In astronomy, quasi-periodic oscillations (QPO) are observed in accretion disks of massive neutron stars or black holes, and it has been suggested that the QPOs arise from the coupling of two oscillatory modes~\cite{Abramowicz03,Kluzniak05,Horak05,Horakthese05}. A particle in the accretion disks  may have radial, vertical and azimuthal epicyclic oscillations around its stable orbit, and because of angular momentum conservation only two of them are independent, and may exhibit a resonance~\cite{Horak05}. Such resonances are called \emph{autoparametric} because there is no external forcing, and the behavior of the resonant mode mimics two well known characteristics of parametric resonance. The amplitude of the resonant mode is a function of time that exhibits an initial exponential increase, and that evolves with a characteristic timescale much larger than the period of the forcing oscillations.

When energy is supplied by an external source, in such a way that a characteristic parameter of the oscillator becomes time-dependant, such as \emph{e.g.} the length of a pendulum, the oscillator response may exhibit parametric resonance~\cite{Landau66}. A paradigm of the parametric resonance is the nonlinear Mathieu equation~\cite{Mond93,Fauve94,Kidachi97,Zounes02,Misbah17,Safonov19}.  A typical feature of parametric resonance is the very slow increase of the parametrically amplified oscillatory mode. There are thus two time scales, a fast one which is the frequency of the forcing and a slow one which characterises the slowly varying amplitude of the forced mode. This strongly suggest the use of a multiscale perturbative expansion to get the oscillators motions, introducing  the two time scales from the start~\cite{Nayfeh73}. This is our analytic tool, which provides  the slow modulation of the oscillators amplitude and phase from the complex amplitude equations  that ensures the consistency of the asymptotic expansion.

For the sake of later comparison, we first recall some results about a parametrically forced oscillator described by the nonlinear Mathieu equation~\cite{Mond93,Kidachi97,Zounes02,Safonov19} without dissipation. This is done in Section~\ref{sec:mathieu} for the parametrically forced Duffing equation [see Eqn.~\eqref{eq:pendulum} of Sec.~\ref{sec:duffing}]. We exhibit the relevant amplitude equation in Sec.~\ref{sec:mathieuamplitude}, and discuss carefully the energy transfer between the oscillator and the external source that provides the parametric excitation in Sec.~\ref{sec:mathieuenergy}.

Our main subject is a nonlinear conservative system with two degrees of freedom. We study  a dimer, made of two point particles that interact with a potential $U_{int}(r)$ where $r$ is the distance between the particles, submitted to an external periodic potential and moving on a line~\cite{FuscoEPJB03}. We take a sinusoïdal external periodic potential of period $a$,
\beq
U_{ext}(x_1,x_2) = U_0\left(2 - \cos \frac{2 \pi x_1}{a} - \cos \frac{2 \pi x_2}{a}\right),
\label{eq:potper}
\eeq
where $x_i$ is the spatial coordinate of the $i$-th particle ($i = 1,2$) and $2 U_0$ is the potential barrier per particle. We consider the commensurate configuration for which the period  $a$ is also the dimer equilibrium length.

This is the subject of our Section~\ref{sec:autoparametric} where we show that there may be an autoparametric behavior of this system in two configurations. In the first one, the oscillator remains trapped in a well of the external potential, and may exhibit autoparametric resonance if the stiffness of the interaction is small enough in a sense that is made quantitative in Section~\ref{sec:well}. The second configuration corresponds to a motion such that the initial kinetic energy is high enough for the dimer center of mass to slide on the external potential and is studied in Section~\ref{sec:sliding}. Three appendices are devoted to technical details. In Section~\ref{sec:conclusion} we sum up our conclusions.

\section{Nonlinear Mathieu equation}
\label{sec:mathieu}

Before addressing our autoparametric oscillator, we consider in this section a nonlinear oscillator with parametric forcing that is described by a nonlinear Mathieu equation. For the sake of later comparison, we focus our discussion on a dissipationless system.

\subsection{A parametrically forced Duffing equation}
\label{sec:duffing}

In order to recall essential features of the parametric resonance, let us first consider the linear Mathieu equation. In dimensionless time units such that the forcing frequency is $2$, it reads
\beq
\frac{\dd^2 u}{\dd t^2} + \left[n^2 + \delta - \epsilon \cos(2 t)\right]u = 0,
\label{eq:mathieu}
\eeq
where $n$ is an integer. In this equation $u$ is the deviation of the oscillator from its equilibrium position, $n^2+\delta$ is the square of its natural frequency and $\epsilon$ is the strength of the forcing. All these parameters are constant with time, and the essential feature of Mathieu equation is that the forcing occurs as a multiplicative term.

Even for a very small forcing $\vert \epsilon\vert \ll 1$, the oscillator may be unstable with an amplitude  that increases exponentially. This instability happens when the parameter $\delta$, which is called the \emph{detuning} is inside a resonance tongue $\delta_n(\epsilon)$ which may be calculated perturbatively~\cite{Landau66,Nayfeh73}. For instance, the main resonance, with the largest possible detuning, occurs for $n = 1$ with $\delta_1(\epsilon) = {\cal O}(\epsilon)$, $-\epsilon/2 <\delta_1< \epsilon/2$. For $n > 1$ the detuning has to be at least an order of magnitude smaller, $\delta_n(\epsilon) = {\cal O}(\epsilon^2)$. Outside the resonant tongue, the displacement $u$ undergoes constant amplitude oscillations.  This well known phenomena is called \emph{parametric resonance}~\cite{Landau66}.

Since Mathieu equation is linear, when the detuning is inside the resonance tongue there is no limiting mechanism so that eventually the amplitude diverges exponentially. This unphysical behavior is avoided when nonlinearities are taken into account. Large amplitude oscillations induce a nonlinear change of the oscillator frequency. This nonlinear detuning makes the energy transfer from the source to the oscillator inefficient, so that eventually the amplitude of the oscillations remains finite.

A relevant physical system is the parametrically forced Duffing equation, which reads, with an appropriate choice of the amplitude unit,
\beq
\frac{\dd^2 u}{\dd t^2} + \left[1 + \delta - \epsilon \cos(2 t)\right]u - u^3 = 0.
\label{eq:pendulum}
\eeq
This system has ben studied quite at length in Ref.~\cite{Kidachi97}. In the remaining of this section, we recall their main results,  adding some complementary results and discussions.

\subsection{Amplitude equation}
\label{sec:mathieuamplitude}

Following the authors of Ref.~\cite{Kidachi97}, we use a multiple scales perturbative expansion that assumes
\beq
u = \sqrt{\epsilon}\left(u_0 + \epsilon u_1 + \ldots\right), \quad \delta = \epsilon \delta_1 + \ldots, \quad \frac{\dd}{\dd t} = \frac{\partial}{\partial T_0} + \epsilon \frac{\partial}{\partial T_1} + \ldots, 
\label{eq:devechmul}
\eeq
where $\ldots$ represents higher order terms, where $T_0 = t$ and where $T_1 = \epsilon t$ is a slow time scale. 

The lowest order solution of~\eqref{eq:pendulum} is easily found to be
\beq
u_0(t) = A(T_1)e^{i T_0} + \overline{A}(T_1)e^{-i T_0},
\label{eq:solu0pendulum}
\eeq
where $A(T_1)$ is a slowly varying complex amplitude and $\overline{A}(T_1)$ its complex conjugate, so that $u_0(t)$ is a real function.

The multiple scale method is well described in textbooks~\cite{Nayfeh73}. At order $\epsilon^{3/2}$, the equations for $u_1(T_0)$ is that of a forced harmonic oscillator. To ensure the consistency of the asymptotic expansion, one requires the vanishing of the \emph{secular terms} which would induce a forcing at the characteristic oscillator frequency. The relevant amplitude equation, which gives the evolution of the oscillator amplitude with the slow time-scale is obtained when the secular terms are eliminated in the equation for $u_1$, and reads
\beq
2 i \frac{\partial A}{\partial T_1} = - \delta_1 A + \frac{1}{2}\overline{A} + 3\vert A \vert^2 A.
\label{eq:amplitudeparametric}
\eeq
There is an obvious fixed point $A = 0$. Setting $\delta A \equiv \delta A_r + i \delta A_i$ where $\delta A_r $ and $ \delta A_i $ are small real perturbations, a linear perturbation analysis gives
\beq
\delta \ddot{A}_r = -\left(\delta_1^2 - \frac{1}{4}\right)\frac{\delta A_r}{4},
\label{eq:stabAzero}
\eeq
showing that the fixed point is a saddle point for $-1/2 \leqslant \delta_1 \leqslant 1/2$, which indicates the parametric instability. The  inequality for the detuning $\delta_1$ is  the lowest order expansion of the instability tongue for the parametric resonance~\cite{Landau66,Nayfeh73}. 

The amplitude equation~\eqref{eq:amplitudeparametric} has been derived in~\cite{Kidachi97} for the dissipationless system~\eqref{eq:pendulum}. The occurence of the complex conjugate term $\overline{A}$ is a typical feature of the nonlinear parametric oscillator, and traces back to the parametric forcing itself~\cite{Fauve94,Misbah17}. Setting $A = R e^{i\phi}$, where the amplitude $R(T_1)$ and the phase $\phi(T_1)$  are real functions of the slow timescale $T_1$, and separating the real and imaginary parts, we obtain the set of equations 
\beq
\begin{cases} \dfrac{\partial R}{\partial T_1} = - \dfrac{R}{4} \sin 2 \phi, \\[2ex] \dfrac{\partial \phi}{\partial T_1} = \dfrac{\delta_1}{2} - \dfrac{1}{4} \cos 2 \phi - \dfrac{3}{2}R^2.\end{cases}
\label{eq:equaparametric}
\eeq
The equation for the phase evidences that the nonlinear effective detuning ${\delta_1} - {3}R^2$ decreases with the amplitude $R$. This simple observation  helps to interpret physically the phase portraits. 

\begin{figure}[htb]
\center
\includegraphics[width=0.32\textwidth]{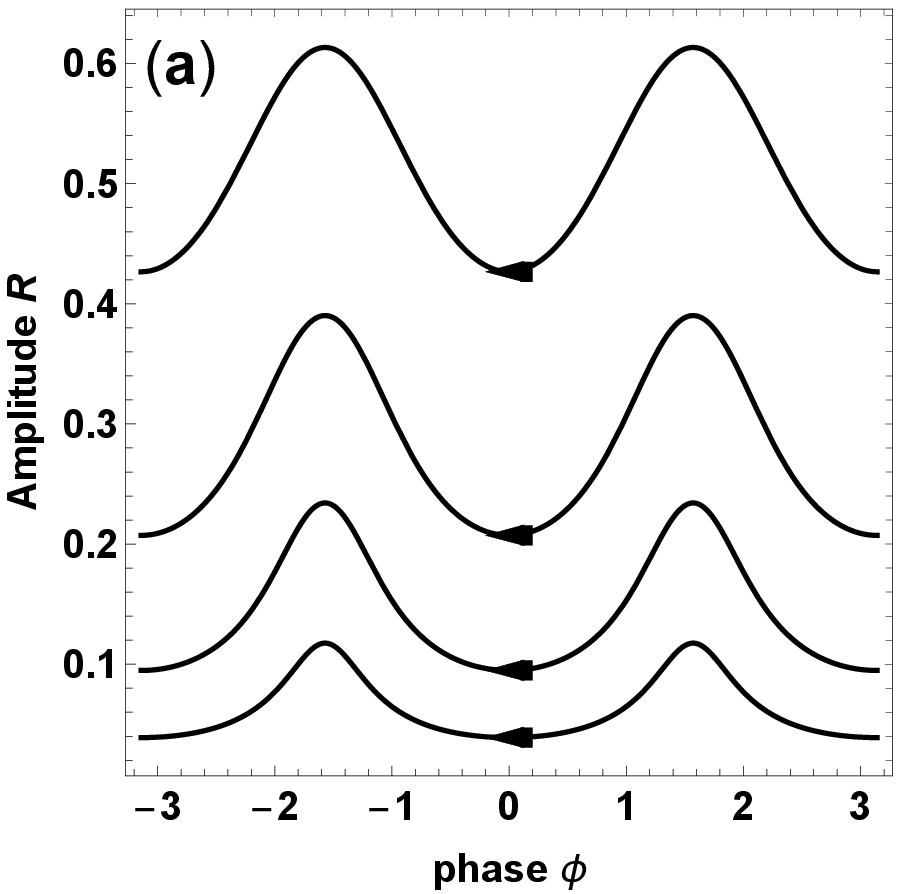}\includegraphics[width=0.32\textwidth]{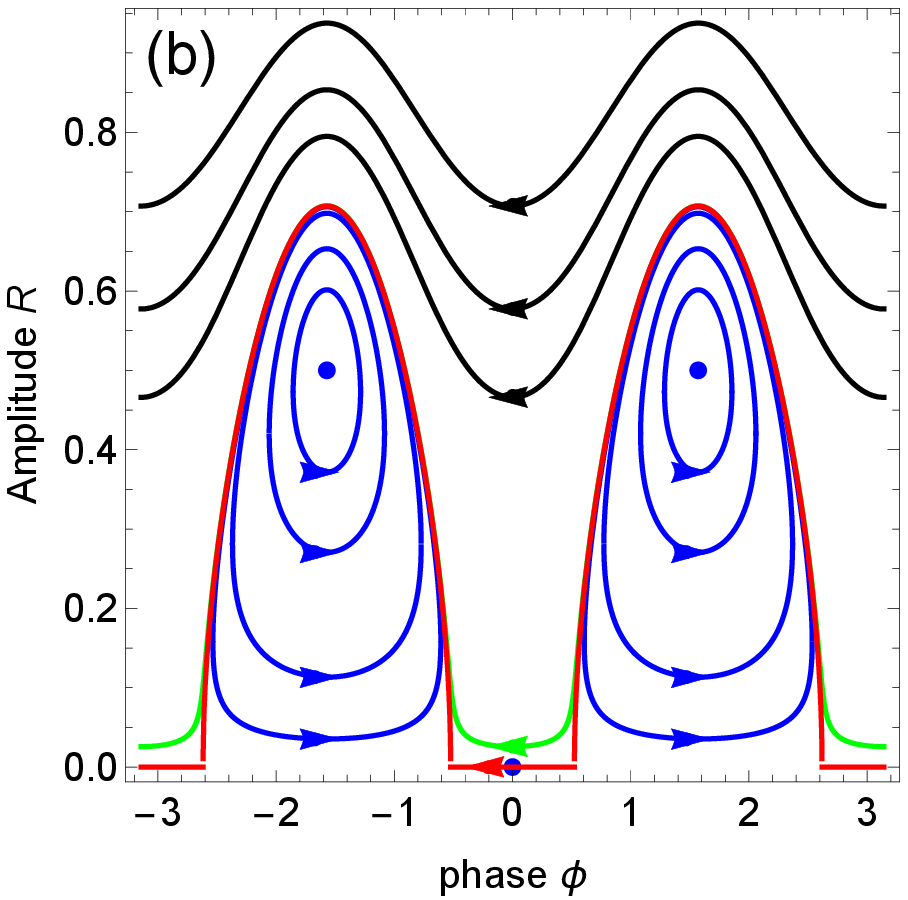}\includegraphics[width=0.32\textwidth]{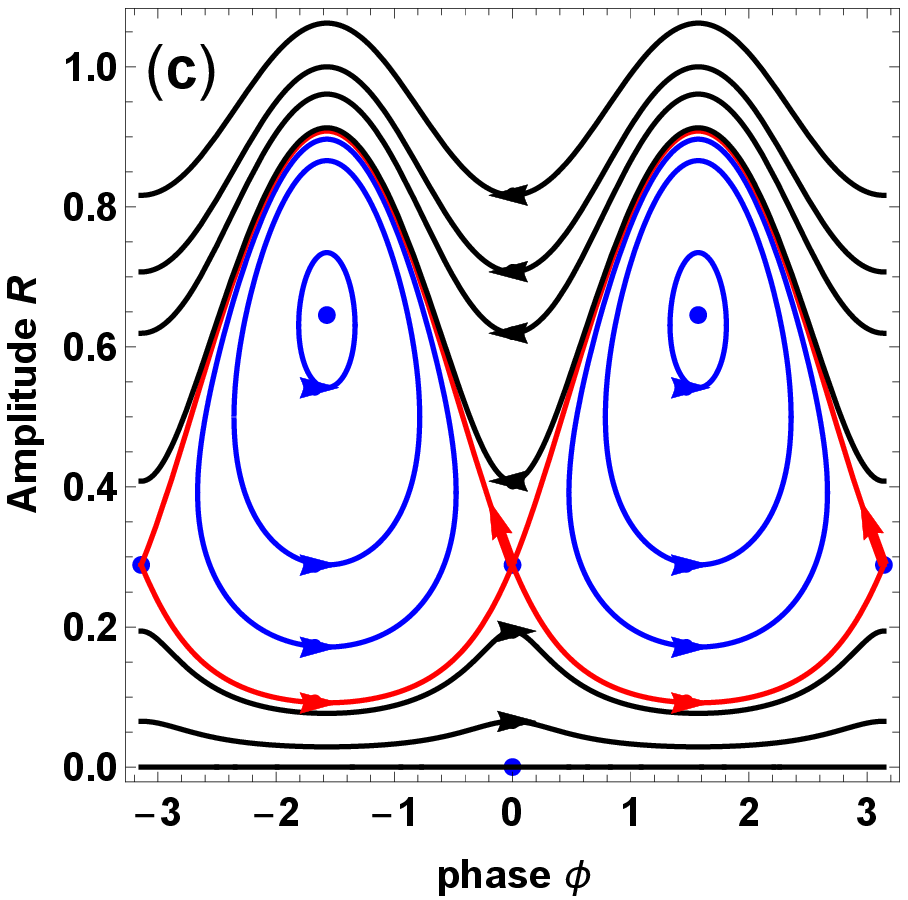}
\caption{\label{fig:portraitmathieu} (Color online) Phase portraits for the amplitude equation~\eqref{eq:equaparametric}. The ordinate is the amplitude $R$ and the abscissa the phase $\phi\in[-\pi,\pi]$. The detuning is (a) $\delta_1 = -0.60$, (b) $\delta_1 = 0.25$ and (c) $\delta_1 = 0.75$. The trajectories in the phase space are given by~\eqref{eq:portraitparametric} for several values of the constant $H$, and the arrows indicate their travel direction for increasing time. See text for details.}
\end{figure}

In order to get the phase portrait of the dynamical system~\eqref{eq:equaparametric}, we search for a constant of the motion. The amplitude equation~\eqref{eq:amplitudeparametric} may be derived from the lagrangian~\cite{Goldstein80}
\beq
L =  i \left(\overline{A} \frac{\partial A}{\partial T_1}  - A \frac{\partial \overline{A}}{\partial T_1}\right) +  \delta_1 \vert A\vert^2 - \frac{1}{4}\left(A^2 + \overline{A}^2\right) - \frac{3}{2}\vert A \vert^4.
\label{eq:lagrangianparametric}
\eeq
Since this lagrangian is time-independent, we readily deduce the conserved quantity
\beq
H =  \frac{3}{2}\vert A \vert^4 +  \frac{1}{4}\left(A^2 + \overline{A}^2\right) - \delta_1 \vert A\vert^2 = R^2\left(\frac{3}{2}R^2 +  \frac{1}{2}\cos 2\phi - \delta_1\right).
\label{eq:hamiltonianparametric}
\eeq
The dynamical system has two degrees of freedom and a constant of the motion so that it is therefore integrable. For a given detuning $\delta_1$, the trajectories in the phase space $(R,\phi)$ are the contour lines with constant $H$ of the function
 \beq
R_\pm^2(\phi) = \frac{1}{3} \left\{\delta_1 - \frac{\cos2 \phi}{2} \pm\left[\left(\delta_1 - \frac{\cos2 \phi}{2}\right)^2 + 6 H\right]^{1/2}\right\}.
\label{eq:portraitparametric}
\eeq
We see from~\eqref{eq:equaparametric} that 
\beq
\frac{\partial \phi}{\partial T_1} = \mp \frac{1}{2}\left[\left(\delta_1 - \frac{\cos2 \phi}{2}\right)^2 + 6 H\right]^{1/2},
\label{eq:derivphi}
\eeq
so that $\partial \phi/\partial T_1 \leqslant 0$ on a branch $R_+(\phi)$, and that $\partial \phi/\partial T_1 \geqslant 0$ on a branch $R_-(\phi)$. The phase portraits are displayed in Fig.~\ref{fig:portraitmathieu}~(a--c) for several values of the detuning $\delta_1$.

\begin{figure}[htb]
\center
\includegraphics[width=0.45\textwidth]{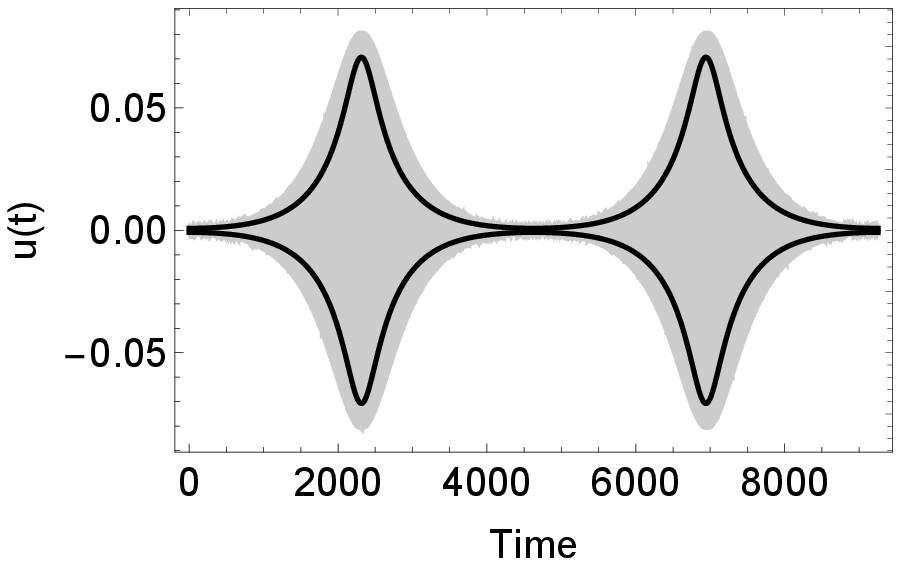}\includegraphics[width=0.45\textwidth]{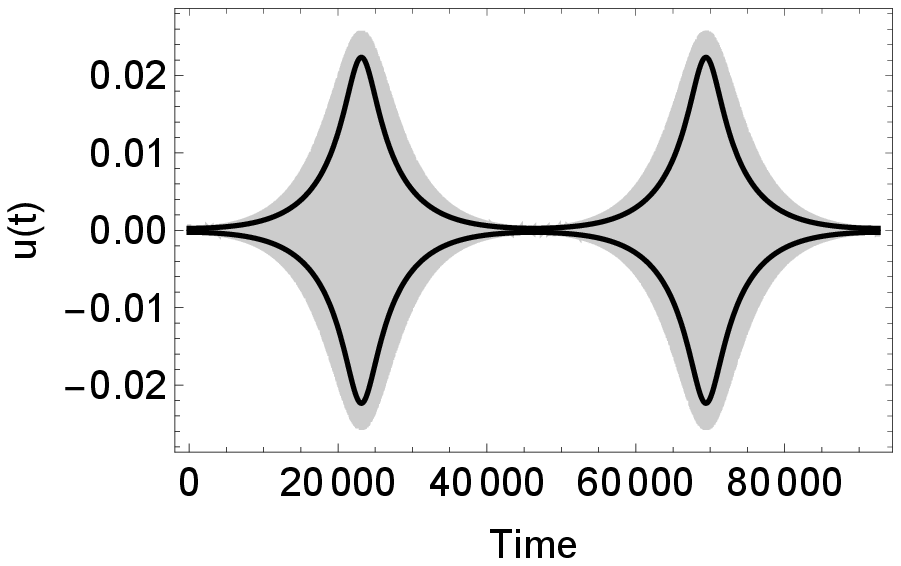}

\caption{\label{fig:compaMathieu} (Color online) Plot of the dimensionless signal $u(t)$ as a function of the dimensionless time $t$ from a numerical integration of the nonlinear Mathieu equation~\eqref{eq:pendulum} (solid gray line; the resolution prevents from distinguishing the quick oscillations) compared to its slowly varying amplitude given by a numerical integration of the system~\eqref{eq:equaparametric} (thick black solid line). The parameters are $\delta_1 = 0.25$, $H = 10^{-4}/6$, $\epsilon = 0.01$ (left) and $\epsilon = 0.001$ (right). There is a striking similarity between the two plots. Paying attention to the axis graduations evidences that the amplitude scales as $\sqrt{\epsilon}$ and that the time scales as $\epsilon^{-1}$.}
\end{figure}

The fixed point $A = 0$, for $H = 0$,  exists regardless of the value of $\delta_1$. It is a saddle when $-1/2 < \delta_1 < 1/2$, and a node otherwise. When $\delta_1<-1/2$ [see Fig.~\ref{fig:portraitmathieu}~(a)], there is no other fixed point, all trajectories in phase space are given by $R_+(\phi)$ for $H >0$ and are obviously open trajectories such that $\phi \in [-\pi,\pi]$. For $-1/2 < \delta_1 < 1/2$ [see Fig.~\ref{fig:portraitmathieu}~(b)], there are two additional nodal points $[\sqrt{(\delta_1+1/2)/3},\pi/2]$ and $[\sqrt{(\delta_1+1/2)/3},3\pi/2]$ for for $H = H_{NP} \equiv -(\delta_1 + 1/2)^2/6$. Closed trajectories ($\phi \in [-\phi^*,\phi^*]$ with $\phi^* < \pi$, solid blue lines) are obtained for $H_{NP} < H < 0$. They are given by $R_+(\phi)$ and $R_-(\phi)$. The separatrix includes the point $A = 0$, corresponds to $H = 0$ and is given by $R_+(\phi)$ (solid red line). Open trajectories (solid black lines) correspond to $0 < H$, and are given by $R_+(\phi)$. The trajectory plotted as a solide green line in Fig.~\ref{fig:portraitmathieu}~(b) is a peculiar open trajectory that is discussed below (see Fig.~\ref{fig:compaMathieu} and Fig.~\ref{fig:energtrans} in Sec.~\ref{sec:mathieuenergy}). Then for $\delta_1 > 1/2$ [see Fig.~\ref{fig:portraitmathieu}~(c)], there are two additional saddle points $[\sqrt{(\delta_1-1/2)/3},0]$ and $[\sqrt{(\delta_1-1/2)/3},\pi]$ for $H = H_{SP} \equiv -(\delta_1 - 1/2)^2/6$. The closed trajectories (solid blue lines) correspond to $H_{NP} < H < H_{SP}$ and the open trajectories (solid black lines) correspond to $H_{SP} < H$. The two saddle points are part of the separatrix (solid red line) which corresponds to $H = H_{SP}$. The upper branch of the separatrix is given by $R_+(\phi)$, the lower one by $R_-(\phi)$. Above the upper separatrix, the open trajectories are given by $R_+(\phi)$, inside the two separatrix branches the closed trajectories are given by $R_+(\phi)$ and $R_-(\phi)$ and below the lower branch the open trajectories are given by $R_-(\phi)$.

It is interesting to get a physical interpretation of the fixed points, and of the topological differences between the phase portraits outside the parametric resonance tongue [compare Fig.~\ref{fig:portraitmathieu}~(a) to Fig.~\ref{fig:portraitmathieu}~(c)]. As seen from~\eqref{eq:equaparametric}, a finite amplitude induces an effective nonlinear detuning $\delta_1- 3 R^2/2 < \delta_1$. In the case of  Fig.~\ref{fig:portraitmathieu}~(c), the detuning is $\delta_1 > 1/2$. The saddle points are such that $R^2=(\delta_1-1/2)/3$,  which gives an effective detuning of $1/2$, that is the upper limit of the instability tongue.  At the nodal points $R^2 = (\delta_1+1/2)/3$, so that the effective detuning is $-1/2$, the lower limit of the instability tongue. If the initial energy of the oscillator is high enough for these nonlinear corrections to take place, its motion will be amplified by the forcing when the effective detuning is in the parametric instability tongue. Strictly speaking, this is not a parametric instability, since for this effect to takes place a finite initial amplitude is required, but it is nevertheless a reminiscence of this instability. In contrast, in the case of  Fig.~\ref{fig:portraitmathieu}~(a), the detuning is $\delta_1 < -1/2$. Since the nonlinear correction to the effective detuning is negative, the effective detuning is always outside the parametrically unstable tongue and the amplification of the relative motion by the center of mass motion is much less than in the previous case.
\smallskip

In the remainder of this section, we consider the parametrically resonant case, $-1/2 < \delta_1 < 1/2$. In Fig.~\ref{fig:compaMathieu}, we compare the numerical solution $u(t)$ of the system~\eqref{eq:pendulum} to its slowly varying amplitude $2 \sqrt{\epsilon}R(\epsilon t)$ given by the amplitude equation~\eqref{eq:equaparametric},  for a detuning $\delta_1 = 0.25$. Since we want to emphasize the parametric amplification, we chose  initial conditions for the amplitude  that are very close to the saddle point in the phase space. Indeed we take $H = 10^{-4}/6$, setting $\phi(t = 0) = 0$ thus $R(t = 0) = 8.16\; 10^{-3}$. The relevant trajectory in phase space is the green solid line in Fig.~\ref{fig:portraitmathieu}~(b), which follows very closely the separatrix. The parametric amplification leads to  a value $R_{max} \approx \sqrt{(2\delta_1+1)/3}$, which is  the maximum amplitude on the separatrix. For consistency, the system~\eqref{eq:pendulum} is numerically integrated with the initial condition $u(t = 0) = 2 \sqrt{\epsilon} R(t = 0)$. The comparison is done in Fig.~\ref{fig:compaMathieu}  for $\epsilon = 0.01$ and $\epsilon = 0.001$. In both case, the slowly varying amplitude of the oscillator is well described by the amplitude equation. Moreover, the similarity of the two plots evidences that $\epsilon$ is indeed a scaling variable, with an oscillator amplitude that scales as $\sqrt{\epsilon}$ and with a slow time-scale that scales as $\epsilon t$, in agreement with the multiple scales expansion.

\begin{figure}[htb]
\center
\includegraphics[width=0.7\textwidth]{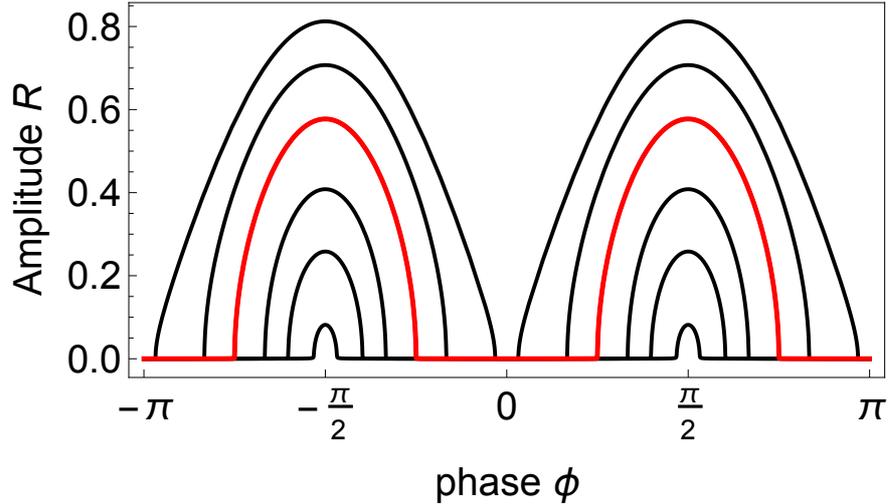}

\caption{\label{fig:Tonguenonlin} Plot of the phase space trajectories $(R(\phi),\phi)$ that include the point $(R_0 = 10^{-4},\phi_0 = 0)$ and for $\delta_1 \in \{0.49, 0.25, 0, -0.25, -0.4, -0.49\}$. For increasing time, they are travelled in the sense of decreasing $\phi$ (hence from right to left). The maximum of the amplitude, $R(\pm\pi/2)$ is an increasing function of $\delta_1$ when $\delta_1 \in [-1/2,1/2]$. The trajectory for $\delta_1 = 0$ is plotted in red to emphasize the shape of the trajectories. }
\end{figure}

For the linear Mathieu equation~\eqref{eq:mathieu}, the growth rate of the oscillations is maximum in the center of the parametric resonance tongue~\cite{Whittacker20}, thus for $\delta_1 = 0$. In Fig.~\ref{fig:Tonguenonlin}, we plot phase space trajectories $R_+(\phi)$ for several values of $\delta_1$ taken in the resonance tongue, such that $-1/2 \leqslant \delta_1 \leqslant 1/2$. To emphasize the parametric amplification, all trajectories include the point $(R_0 = 10^{-4},\phi_0 = 0)$, which corresponds to a very small value of $H$, since $H$ is of order $R_0^2$ as seen from Eqn.~\eqref{eq:hamiltonianparametric}. The corresponding open trajectories are thus very close to the separatrix defined by $H = 0$, so that the maximum amplitude is very close to $R_{max}$. Therefore, the maximum amplitude of a parametrically unstable trajectory decreases when $\delta_1$ decreases while remaining inside the parametric resonance tongue. This behavior is characteristic of the nonlinear Mathieu equation~\eqref{eq:pendulum}. Physically, because of the effective nonlinear detuning, the system will cross the parametric instability tongue from $\delta_1$ down to $-1/2$. Therefore, the smaller the value of $\delta_1$, the smaller the duration of the parametric amplification and the smaller the maximum amplitude of the oscillations.

\subsection{Energy transfer}
\label{sec:mathieuenergy}

The basis phenomenon in parametric resonance is energy transfer from an external source toward the parametrically excited oscillator, when the forcing frequency is in the resonance tongue. For a linear oscillator without dissipation, the energy is permanently transfered to the oscillator, whose amplitude growths without limit no matter how small the amplitude of the parametric forcing. When dissipation is taken into account, the forcing has to be sufficient to overcome the dissipation, but when this condition is fulfilled the amplitude of a linear oscillator once again growths without limit. If the nonlinearities are included in the model, the oscillator amplitude saturates to a finite value, as shown by the bifurcation analysis of the relevant amplitude equations~\cite{Fauve94,Misbah17}. Physically, the change of characteristic frequency with the amplitude of the oscillations induces a nonlinear detuning between the source and the oscillator, and the amplitude saturates when the energy transferred by the source compensates the dissipated energy. 

The case of a non dissipative and nonlinear oscillator is slightly more involved.  In particular, the slow periodic modulation of the oscillator amplitude exhibited in Fig.~\ref{fig:compaMathieu} in the parametrically unstable domain of the detuning $\delta$ is a distinctive feature of the dissipationless Mathieu equation~\cite{Mond93,Kidachi97}. It evidences the conservation of the oscillator energy on large time scale, which means that the energy flow between the source and the oscillator  cannot always be in the same direction, but must be periodically reversed. The aim of this section is to clarify this energy transfer.

The equation~\eqref{eq:pendulum} may be derived from the lagrangian
 \beq
 {\cal L} = \frac{\dot u^2}{2} - \left(1 + \delta - \epsilon \cos 2 t\right)\frac{u^2}{2} + \frac{u^4}{4}.
 \label{eq:lagrangienMathieuNL}
 \eeq
 This lagrangian is explicitely time-dependent, which evidences that the oscillator is not conservative, since the parametric excitation induces an energy transfer between an external source and the oscillator. The corresponding hamiltonian ${\cal H}$ is not a constant of the motion, and using the Lagrange equation~\eqref{eq:pendulum} it is easily seen that $\dd  {\cal H}/\dd t = \epsilon u^2 \sin 2 t$. From~\eqref{eq:solu0pendulum}, we may write $u(t) = 2 R \sqrt{\epsilon}\cos(t + \phi)$. Expanding the trigonometric function, we readily obtain
 \beq
 \frac{\dd {\cal H}}{\dd t} = 2 \epsilon^2 R^2\left[\sin 2 t + \frac{1}{2}\sin(4 t + 2 \phi) - \frac{1}{2}\sin 2 \phi\right].
 \label{eq:derivHamMathieu}
 \eeq
 Let us now average this equation on the quick time scale. Then the amplitude $R$ and phase $\phi$, which both only depend on the slow time scale $\epsilon t$ may be considered as constants. Therefore
 \beq
 \langle  \frac{\dd {\cal H}}{\dd t}  \rangle \equiv \frac{1}{2 \pi}\int_{-\pi}^{\pi}\frac{\dd {\cal H}}{\dd t}  \dd t = - \epsilon^2 R^2 \sin 2 \phi.
\label{eq:derivHamMathieumoy}
 \eeq
This equation exhibits the slow  energy transfer between the oscillator and the external source, the direction of the transfer depending upon the phase $\phi$. Note that, since $R_+(\phi)$ is an even function of the phase, if we average the energy transfer on the \emph{slow} timescale, we get that
  \beq
\langle \langle  \frac{\dd {\cal H}}{\dd t}  \rangle\rangle \equiv \frac{1}{2 \pi}\int_{-\pi}^{\pi}\langle  \frac{\dd {\cal H}}{\dd t}  \rangle  \dd \phi = 0,
\label{eq:derivHamMathieumoy}
 \eeq
 which is consistent with the fact that the amplitude equation~\eqref{eq:amplitudeparametric} may be derived from a time-independant lagrangian~\eqref{eq:lagrangianparametric} that is the signature of a conservative system.

\begin{figure}[htb]
\center
\includegraphics[width=0.7\textwidth]{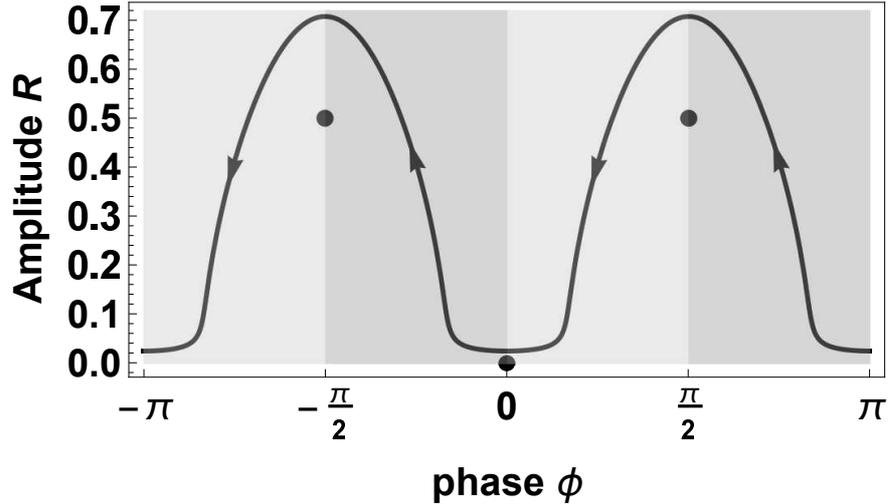}

\caption{\label{fig:energtrans} Phase space trajectory $R_+(\phi)$ in the plane $(\phi ,R)$ for $\delta_1 = 0.25$ and $H = 10^{-3}/6$ [green solid line in Fig.~\ref{fig:portraitmathieu}~(b)]. The arrows indicate the travel direction along the trajectory. In medium gray we indicate the regions of increasing amplitude, for which $\sin 2 \phi <0$ and energy is given to the oscillator by the source. In light gray we indicate the regions of decreasing amplitude, for which $\sin 2 \phi >0$ and energy is returned by the oscillator to the source.}
\end{figure}

A phase space trajectory that corresponds to the parametric instability [solid green line in Fig.~\ref{fig:compaMathieu}--(b)] is shown in Fig.~\ref{fig:energtrans}. In the parametrically unstable region ($-1/2 < \delta < 1/2$) the amplitude is $R_+(\phi)$ given in Eqn.~\eqref{eq:portraitparametric}, and injecting this expression in Eqn.~\eqref{eq:equaparametric} shows that $\partial \phi/\partial T_1 < 0$, which explains the travel direction along the trajectory displayed in Fig.~\ref{fig:energtrans}. To be specific, let us begin with $\phi = \pi$. As time increases, the phase decreases and the amplitude increases up to its maximum which is reached when $\phi = \pi/2$. Physically, this part of the trajectory corresponds to a parametric amplification of the oscillator, which gains energy from the source. This is consistent with Eqn.~\eqref{eq:derivHamMathieumoy} because for $2 \phi \in [\pi,2 \pi)$ we have $\sin 2 \phi <0$ hence an increase in the oscillator energy. In the next quarter period of amplitude modulation, the amplitude decreases, $2\phi \in [0,\pi]$ so that $\sin 2 \phi$ is positive and the oscillator restitutes the energy to the source (which in this case behaves as a well). On the average, the energy is conserved.

\section{Autoparametric resonance}
\label{sec:autoparametric}

In this section, we consider a nonlinear conservative system with two degrees of freedom. More precisely, we study  a dimer, made of two point particles that interact with a potential $U_{int}(r)$ where $r$ is the distance between the particles, submitted to an external periodic potential and moving on a line. In this system, the nonlinear coupling between the center of mass motion and the oscillations around the dimer equilibrium length may exhibit an autoparametric resonance.

The equations of motions read
\beq
\begin{cases} m \ddot x_1 = - \partial U_{int}/\partial x_1 - (2 \pi U_0/a)\sin (2 \pi x_1/a), \\ m \ddot x_2 =  - \partial U_{int}/\partial x_2 - (2 \pi U_0/a)\sin (2 \pi x_2/a), \end{cases}
\label{eq:motiondim}
\eeq
where $m$ is the mass of each particle and where $\ddot x_i \equiv \dd^2 x_i/\dd t^2$. There are two configurations in which such a system may exhibit autoresonance.

\subsection{Dimer in a well.}
\label{sec:well}

Let us first assume that the dimer, during its motions, does not escape the external potential well. Since the system is conservative, this assumption implies a small enough initial energy. Because of the commensurability assumption, when both particles are in adjacent wells of the external potential the interaction potential is also minimum. It is sufficient to take its harmonic approximation with a stiffness $k$, so that the equations of motions read
\beq
\begin{cases} m \ddot x_1 = k(x_2 - x_1 - a) - (2 \pi U_0/a)\sin (2 \pi x_1/a), \\ m \ddot x_2 = k(x_1 - x_2 + a) - (2 \pi U_0/a)\sin (2 \pi x_2/a). \end{cases}
\label{eq:motiondim}
\eeq
We rescale the variables using $a/(2 \pi)$ as the unit length, $U_0$ as the unit energy and therefore $\sqrt{m a^2/(4 \pi^2 U_0)}$ as the unit time. Moreover, we introduce the normal modes as
\beq
x \equiv \frac{x_1 + x_2}{2} - \pi, \quad y \equiv \frac{x_2 - x_1}{2} - \pi,
\label{eq:modes}
\eeq
in dimensionless units. The mode $x$ is the center of mass motion, and the mode $y$ the relative motion of the dimer particles. The potential energy reads
\beq
U(x,y) = 2 \left(1 - \cos x \cos y + K y^2\right),
\label{eq:potadim}
\eeq
where $K = {k a^2}/({4 \pi^2 {U}_0})$ is the dimensionless stiffness. The equations of motion become 
\beq
\begin{cases}\ddot x &= -  \sin x \cos y, \\  \ddot y &= - 2  K y -   \cos x\sin y, \end{cases} 
\label{eq:FuscoExact}
\eeq

\subsubsection{Heuristic approach}
\label{sec:heuristic}

To begin with, we basically follow the heuristic approach of Ref.~\cite{FuscoEPJB03}. We introduce a small parameter $\epsilon$, with $\vert \epsilon \vert \ll 1$ and let $x = \OG(\epsilon)$ and $y = \OG(\epsilon)$, making no assumption on the stiffness $K$. Up to order $\epsilon^3$, the equations of motions are coupled Duffing equations,
\beq
\begin{cases}\ddot x &= -  x + \dfrac{1}{2} y^2 x + \dfrac{1}{6} x^3, \\  \ddot y &= - (2  K +1) y + \dfrac{1}{6} y^3 + \dfrac{1}{2} y x^2. \end{cases} 
\label{eq:FuscoStiff}
\eeq
At leading order, the solutions may be written 
\beq
x(t) = 2\epsilon a \cos(t + \phi), \quad y(t) = 2\epsilon b \cos (\omega t + \psi),
\label{eq:heuristicbase}
\eeq
where $\omega^2 \equiv 1 + 2 K$ and where the amplitudes $a$ and $b$ and the phase $\phi$ and $\psi$ are real functions that for the moment are supposedly slowly varying functions of time. Injecting these expressions in the system~\eqref{eq:FuscoStiff}, we may rewrite it 
\beq
\begin{cases}\ddot x +\left[1 -{b^2 \epsilon^2} - {b^2 \epsilon^2}\cos(2 \omega t + 2 \psi) \right]x - \dfrac{1}{6} x^3 \approx 0, \\[2ex]  \ddot y +\left[\omega^2 - {a^2 \epsilon^2} - {a^2 \epsilon^2}\cos(2 t + 2 \phi) \right]y - \dfrac{1}{6} y^3 \approx 0. \end{cases} 
\label{eq:FuscoStiffApprox}
\eeq
These equations help to get a physical picture, since they are both very similar to the parametric Duffing equation~\eqref{eq:pendulum} studied in Sec.~\ref{sec:mathieu}. In particular, we see that each mode may behave as a parametric forcing of the other one. From our previous calculations, we deduce that a necessary conditions for a parametric resonance to appear is that $\omega^2 = 1 + {\cal O}(\epsilon^2)$, which means that the stiffness of the dimer interaction has to be very small, $K = {\cal O}(\epsilon^2)$. In what follows, we will see that on this condition, the parametric amplification of the relative motion by the center of mass motion may happen, which is called by analogy with Sec.~\ref{sec:mathieu} an \emph{autoparametric resonance}~\cite{Horak05}. The case of the dimer in a well with a strong bond is studied in Ref.~\cite{Maddi22papier2}.

\subsubsection{Amplitude equations}
\label{sec:wellcalc}

A consistent description of the nonlinear oscillations of the dimer may be obtained from a systematic asymptotic expansion. Formally, we search a solution of the system~\eqref{eq:FuscoExact} perturbatively, using the method of multiple scales. We introduce  successive times scales $T_0 = t, T_2 = \epsilon^2 t, \ldots$, and power expansions
\beq
\begin{cases} x(t) = \epsilon X_1(T_0,T_2,\ldots) + \epsilon^3 X_3(T_0,T_2,\ldots) + \ldots, \\ y(t) = \epsilon Y_1(T_0,T_2,\ldots) + \epsilon^3 Y_3(T_0,T_2,\ldots) + \ldots, \\ \dfrac{\dd^2}{\dd t^2} = \dfrac{\partial^2}{\partial T_0^2} + 2\epsilon^2 \dfrac{\partial^2}{\partial T_0 \partial T_2} + \ldots. \\ \end{cases}
\label{eq:defechmul}
\eeq
Up to order $\epsilon^3$, the equations of motions are correctly given by the system~\eqref{eq:FuscoStiff}. As explained in Sec.~\ref{sec:heuristic}, we consider a small stiffness, setting $K \equiv \epsilon^2 \K$ where $\K$ is of order one.

At order $\OG(\epsilon)$, the two modes $X_1(T_0)$ and  $Y_1(T_0)$ behave as free uncoupled harmonic oscillators with the same frequency (1 in our dimensionless variables) so that the solutions at this order are
\beq
X_1 = A(T_2) e^{i T_0} + \overline{A}(T_2) e^{-i T_0}, \quad Y_1 = B(T_2) e^{i  T_0} + \overline{B}(T_2) e^{-i  T_0},
\label{eq:softorder1}
\eeq
where $A(T_2) $ and $B(T_2)$ are slowly varying complex amplitudes.

At order $\epsilon^3$, the equations for $X_3(T_0)$ and  $Y_3(T_0)$ are those for forced harmonic oscillators. As said in Sec.~\ref{sec:mathieuamplitude}, the consistency of the asymptotic expansion requires the vanishing of the secular terms which would induce a forcing at the characteristic oscillator frequency. This solvability condition provides the following coupled amplitude equations~:
\beq
\begin{cases} 2 i  \dfrac{\partial A}{\partial T_2} &= \dfrac{1}{2}\vert A \vert^2 A + \vert B \vert^2 A + \dfrac{1}{2} B^2 \overline{A}, \\[2ex]
2 i  \dfrac{\partial B}{\partial T_2} &= -2 \K B + \dfrac{1}{2}\vert B \vert^2 B + \vert A \vert^2 B+ \dfrac{1}{2} A^2 \overline{B}.\end{cases}
\label{eq:Amplitude}
\eeq

\subsubsection{Simple solutions}
\label{sec:wellparametric}

The system~\eqref{eq:Amplitude} has the obvious solution $(A = 0,B = 0)$, which is marginally stable since the frequency for $A$ vanishes. This solution is basically worthless and reflects the stable equilibrium position of the dimer in the potential well.

Another simple solution is $(A_0 \neq 0,B = 0)$. The equation for $A$ is $4 i(\partial A/\partial T_2) = \vert A \vert^2 A$. Setting $A_0 \equiv a_0 e^{i \phi_0}$, and separating real and imaginary parts in this equation, we get $A_0(T_2) = a_0 \exp-i \left(a_0^2 T_2/4 + \phi_0\right)$, where $a_0$ and $\phi_0$ do not depend on $T_2$. The center of mass motion is thus
\beq
x(t) = 2 \epsilon a_0 \cos\left[\left(1 - \frac{\epsilon^2 a_0^2}{4}\right)t + \phi_0\right],
\label{eq:comRP}
\eeq
which exhibits an amplitude dependent frequency (Borda frequency) because of the nonlinearities. When this solution for $A$ is injected into the equation for $B$, we get
$$
2 i  \dfrac{\partial B}{\partial T_2} = -2 \K B + \dfrac{1}{2}\vert B \vert^2 B + a_0^2 B + \dfrac{a_0^2}{2} e^{-i a_0^2 T_2/2} \overline{B},
$$
where we have chosen $\phi_0 = 0$. Setting $B(T_2) = D(T_2) \exp\left(- i a_0^2  T_2/4\right)$, we get the autonomous equation 
\beq
\frac{\partial D}{\partial T_2} = i \left({\K}- \frac{ a_0^2}{4}\right)D +i \frac{a_0^2}{4}  \overline{D} -\frac{i }{4}\vert D \vert^2 D.
\label{eq:parametricD}
\eeq
We recognise the amplitude equation~\eqref{eq:amplitudeparametric} for the parametric oscillator. This is the only equation relevant to the linear stability analysis of the solution $(A_0,0)$ since the coupling between $A$ and $B$ is at least quadratic in the first equation of~\eqref{eq:Amplitude}. The solution $(A = A_0,B = 0)$ is parametrically unstable for $0 \leqslant \K \leqslant a_0^2/2$. Physically, it means that the center of mass motion may induce a parametric amplification of the relative motions between the two particles of the dimer. Since there is no external source to induce the parametric resonance, we follow the terminology of  Ref.~\cite{Horak05} and call this resonant coupling between the two oscillatory modes of the dimer an autoparametric resonance.

There is also a solution $(A = 0,B \neq 0)$, with $4 i(\partial B/\partial T_2) = -4\K B +\vert B \vert^2 B$. Setting $B \equiv b_0 e^{i \psi_0}$, separating real and imaginary part, we get $B(T_2) = b_0 \exp-i \left[\left(\K - b_0^2 /4\right))T_2 + \psi_0\right]$, where $b_0$ and $\psi_0$ do not depend on $T_2$. To study the linear stability of this solution, let $A = \delta A'$ such that $\vert \delta A' \vert \ll 1$. We get from~\eqref{eq:Amplitude}
$$
2 i  \frac{\partial \delta A'}{\partial T_2} = \frac{ b_0^2}{4}\delta A' + \frac{a b_0^2}{2}e^{2 i \psi} \overline{\delta A'}.
$$
We obtain an autonomous equation by setting $\delta A' = \delta A e^{i \psi}$,
$$
 i  \frac{\partial \delta A}{\partial T_2} =\left({\K} + \frac{b_0^2}{4}\right)\delta A + \frac{ b_0^2}{4}\overline{\delta A}.
 $$
If we write $\delta A = \delta A_r + i \delta A_i$, separating real and imaginary part we get
$$
\frac{\partial \delta A_r}{\partial T_2} = {\K}\delta A_i, \qquad \frac{\partial \delta A_i}{\partial T_2} = -\left( {\K} + \frac{ b_0^2}{2} \right)\delta A_r.
$$
Setting $\delta A_r = \delta A_r^0 e^{\sigma T_2}$ and $\delta A_i = \delta A_i^0 e^{\sigma T_2}$, we find eventually
$$
\sigma^2 =  - {\K} \left(\frac{ b_0^2}{2} + {\K}\right) \leqslant 0,
$$
which indicates that this solution is always stable. There is therefore no parametric amplification of the center of mass motion by the relative motion of the particles in the dimer.

\subsubsection{Constants of the motion and phase portrait}
\label{sec:wellportrait}

We will now fully describe the solutions of the amplitude equations~\eqref{eq:Amplitude}. Let us express the complex amplitudes $A$ and $B$ as 
\beq
A(T_2) = a(T_2) e^{i \phi(T_2)}, \quad B(T_2) = b(T_2) e^{i \psi(T_2)},
\label{eq:chgtvarequaampl}
\eeq
where the real functions $a$ and $b$ are the amplitudes, and where the real functions $\phi$ et $\psi$ are the phases.  Taking real and imaginary parts in both equations of the system~\eqref{eq:Amplitude}, one gets
\beq
\begin{cases} \dot a = \dfrac{ b^2 a}{4} \sin \theta,  \\[2ex]  \dot b = - \dfrac{ a^2 b}{4} \sin \theta,  \\[2ex] \dot\theta =  2{\K} + \dfrac{1}{2}\left(b^2 -a^2 \right)(1 + \cos \theta),
\end{cases}
\label{eq:equaampldim3}
\eeq
where for simplicity $\dot a \equiv \partial a/\partial T_2$, and the same for $\dot b$. We notice that the phase space of the dynamical system is actually of dimension 3, since the phases are involved only through their difference $\theta \equiv 2(\psi - \phi)$. Indeed, the knowledge of the three functions $a$, $b$ and $\theta$ is sufficient to get all dynamical variables since
\beq
\phi(t) = -\int_0^t \left[\dfrac{ a(u)^2}{4} + \frac{ b(u)^2 }{2} + \dfrac{ b(u)^2 }{4}\cos \theta(u)\right]\dd u, \qquad \psi(t) = \phi(t) + \theta(t)/2.
\label{eq:phases}
\eeq

In order to get a full phase portrait of our system, it is convenient to search for constants of the motion for the system~\eqref{eq:equaampldim3}. There is an  obvious one,
\beq
a^2 + b^2 \equiv N \Longleftrightarrow a \dd a = -b \dd b.
\label{eq:norme}
\eeq
There is a second independent constant which may be found as in Ref.~\cite{Horak05}. Let us consider $\theta$ as a function of $a$, so that 
$$
\dot \theta = \frac{\dd \theta}{\dd a} \dot a = \dfrac{ b^2 a}{4} \sin \theta \frac{\dd \theta}{\dd a},
$$
where we used the first equation of~\eqref{eq:equaampldim3}. Injecting this result in the equation for $\dot \theta$, and multiplying all terms by $a \dd a$, we get 
\beqa
\dfrac{ b^2 a^2}{4} \sin \theta \dd \theta - \dfrac{ b^2}{2} \cos \theta a \dd a + \dfrac{ a^2}{2} \cos \theta a \dd a -  2\K a \dd a +  \dfrac{ a^3}{2}  \dd a -  \dfrac{ b^2}{2}  a \dd a = 0 \nonumber \\ - \dfrac{ b^2 a^2}{4} \dd \left(\cos \theta\right) - \dfrac{ b^2}{2} \cos \theta \dd \left(\frac{a^2}{2}\right) -  \dfrac{ a^2}{2} \cos  \theta \dd \left(\frac{b^2}{2}\right) - \K \dd\left({a^2}\right) + \dd \left(\frac{a^4}{8}\right) + \dd \left(\frac{b^4}{8}\right) = 0 \nonumber 
\eeqa
where in the second line we used~\eqref{eq:norme}. We thus get a second constant of motion, 
\beq
J \equiv \frac{a^4}{8} + \frac{b^4}{8} - \K a^2 - \dfrac{ b^2 a^2}{4} \cos  \theta,
\label{eq:Horak}
\eeq
which is obviously independent on the first one. In appendix~\ref{App:Noether}, we derive both constants using lagrangian formalism and Noether theorem. Because of these two independent constants of the motion, our system is integrable. 

Let us introduce the dynamical variable $\chi(t)$, such that $0 \leqslant \chi \leqslant 1$, as
\beq
a(t)^2 \equiv N [1 - \chi(t)], \quad b(t)^2  = N\chi(t).
\label{eq:defxi}
\eeq
We show in appendix~\ref{App:constanteN} that an appropriate definition of the small parameter $\epsilon$ allows to take $N = 1$ without any loss in generality, and we will do it henceforward.

Using~\eqref{eq:equaampldim3}, we get the dynamical equations
\beq
\begin{cases} \dot\chi &= -\dfrac{1}{2} \chi(1 - \chi )\sin  \theta, \\[2ex] \dot \theta &= 2\K + \dfrac{1}{2}(2\chi - 1)(1 + \cos  \theta).\end{cases}
\label{eq:equaamplchi}
\eeq
There is a pair of fixed points 
\beq
\chi^* = 0, \quad \cos  \theta^* = {4\K}-1,
\label{eq:fpsepa}
\eeq
if we are in the autoparametric (AP; $0 \leqslant \K \leqslant 1/2$) regime. Since the constant of the motion is
\beq
 J(\chi,\theta) = \frac{1}{8} - \frac{1}{4}\chi(1 - \chi)(1 + \cos \theta) -  \K (1 - \chi),
\label{eq:valJ}
\eeq
these fixed points correspond to $J = 1/8-\K$. Introducing small perturbations $(\delta\chi, \delta\theta)$, the linear stability analysis gives
\beq
{\delta \dot\chi} = - \frac{1}{2}\sin \theta^* \delta \chi, \quad {\delta \dot\theta} = \sin \theta^* \delta \theta,
\label{eq:stabfpsepa}
\eeq
which shows that these fixed points are saddle points. The other fixed point is
\beq
\chi^* = \frac{1}{2} -{\K}, \quad  \theta^* = 0,
\label{eq:fpnode}
\eeq
which only exists in the AP regime, and which corresponds to $J = -\K(1 + \K)/2$. The linear stability analysis gives
$$
{\delta\ddot\chi} = -\left(\frac{1}{2} - {\K}\right)\left(\frac{1}{2} + {\K}\right)\delta \chi,
$$
showing that this fixed point is a node.

Knowing the constant of motion $J$, we can plot the phase portrait of the dynamical system in the plane $(\theta,\chi)$. In the AP regime, $J$ has the local minimum $J = 1/8 - \K$ for the fixed points~\eqref{eq:fpsepa}, and the absolute minimum $J = -{\K}\left(\K + 1\right)/2$ for the fixed point~\eqref{eq:fpnode}. Therefore
\beq
\begin{cases}
-\K(\K+ 1)/2 \leqslant J \leqslant 1/8 \qquad \hbox{(AP)}, \\ 1/8 - \K \leqslant J \leqslant 1/8 \qquad \hbox{(NR)}.
\end{cases}
\label{eq:varJ}
\eeq

The available range of $\chi$ is obtained by taking the square of the equation for $\dot\chi$ in~\eqref{eq:equaamplchi}, expressing the result as a function of $\cos \theta$ and eliminating  $\theta$ with the help of~\eqref{eq:Horak}. Indeed, we eventually get
\beq
{4}\left(\frac{\dd \chi}{\dd t}\right)^2 = \chi^2 (1 - \chi)^2 - \left[\frac{\chi^2 + (1 - \chi)^2}{2} - {4 \K} (1-\chi) - {4 J}\right]^2 \equiv F(\chi)^2 - G(\chi)^2,
\label{eq:Horakxi}
\eeq
which shows that the motion is restricted to those values of $\chi$ for which the right-hand-side is positive. The solutions of $F(\chi) = + G(\chi)$ are
\beq
\chi_{\pm}^+=  \frac{1}{2} -{\K} \pm \sqrt{\left({\K} + \frac{1}{2}\right)^2 - \frac{1}{4} + {2 J}}.
\label{eq:solchiplusmoins}
\eeq
Since $\chi \in [0,1]$ by construction, we must have $\chi_{-}^+ \geqslant 0$, which implies $J > 1/8 - \K$. Otherwise we must take the solution of $F(\chi) =  - G(\chi)$, that reads
\beq
\chi_{+}^- =1 - \frac{1}{\K}\left(\frac{1}{8} - {J}\right),
\label{eq:solchimoinsplus}
\eeq

Using Eqn.~\eqref{eq:valJ}, we see that the trajectories in phase space are given by
\beq
\cos \theta = \frac{\chi^2 + (1 - \chi)^2 - 8 \K (1-\chi) - 8 J}{2 \chi(1 - \chi)}  = \frac{G(\chi)}{F(\chi)},
\label{eq:equaphase}
\eeq
Searching for extremal values of $\cos \theta$, we solve $G' F - F ' G = 0$ and get
\beq
\chi^*_\pm = 1 - \frac{1 - 8J}{8  \K} \left(1 \pm \sqrt{1 - \frac{8  \K}{1 - 8J}}\right).
\label{eq:chistar}
\eeq
There are no real solutions in the NR case, which means that $\theta \in [-\pi,\pi]$ for all possible values of $J$ in the NR case, so that only open trajectories occur. 

In the AP case, there is a real solution $0 < \chi^*_- < 1$ for $8 \K < 1 - 8 J$ (the solution $\chi^*_+ < 0$ has to be rejected). In that case, the trajectories are closed curves, such that $\theta \in [-\theta^*_-,\theta^*_-]$ with $\theta^*_- < \pi$. For $\theta = 0$, we deduce from~\eqref{eq:equaphase} that $F(\chi) = G(\chi)$, so that  $\chi(\theta = 0)$ on the closed trajectories is given by $\chi_{\pm}^+$ in~\eqref{eq:solchiplusmoins}. The limit $\chi_-^+ = \chi_+^+$, which requires the vanishing of the square root, is the fixed point~\eqref{eq:fpnode}. These closed trajectories are forbidden for $J \geqslant 1/8 - \K$. The separatrix, between the closed trajectories and the open ones, is thus given by Eqn.~\eqref{eq:equaphase} for $J = 1/8 - \K$, and therefore includes the two saddle points~\eqref{eq:fpsepa}. The linear stability analysis of these points provides the orientation of the separatrix. The orientation of the other trajectories is given by~\eqref{eq:equaamplchi}, which shows that $\theta$ increases with time and that $\chi$ increases  (decreases) with time for $\theta \in [-\pi,0]$ ($\theta \in [0,\pi]$).

\begin{figure}[htb]
\center
\includegraphics[width=0.475\textwidth]{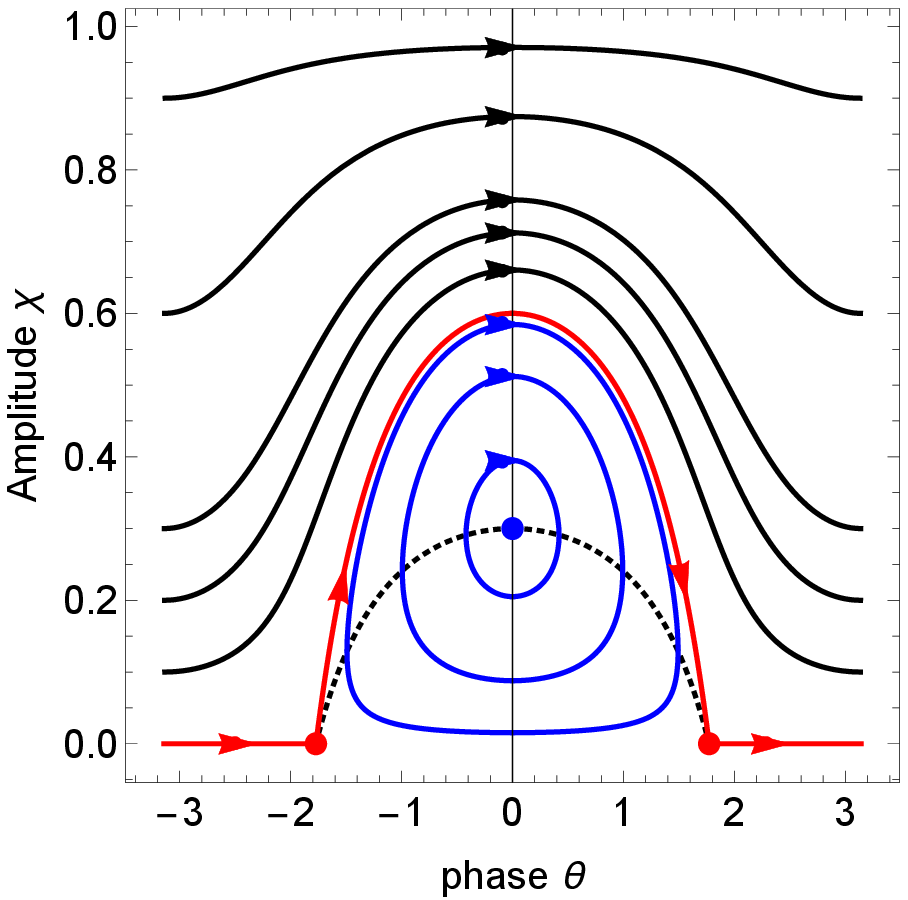} \includegraphics[width=0.475\textwidth]{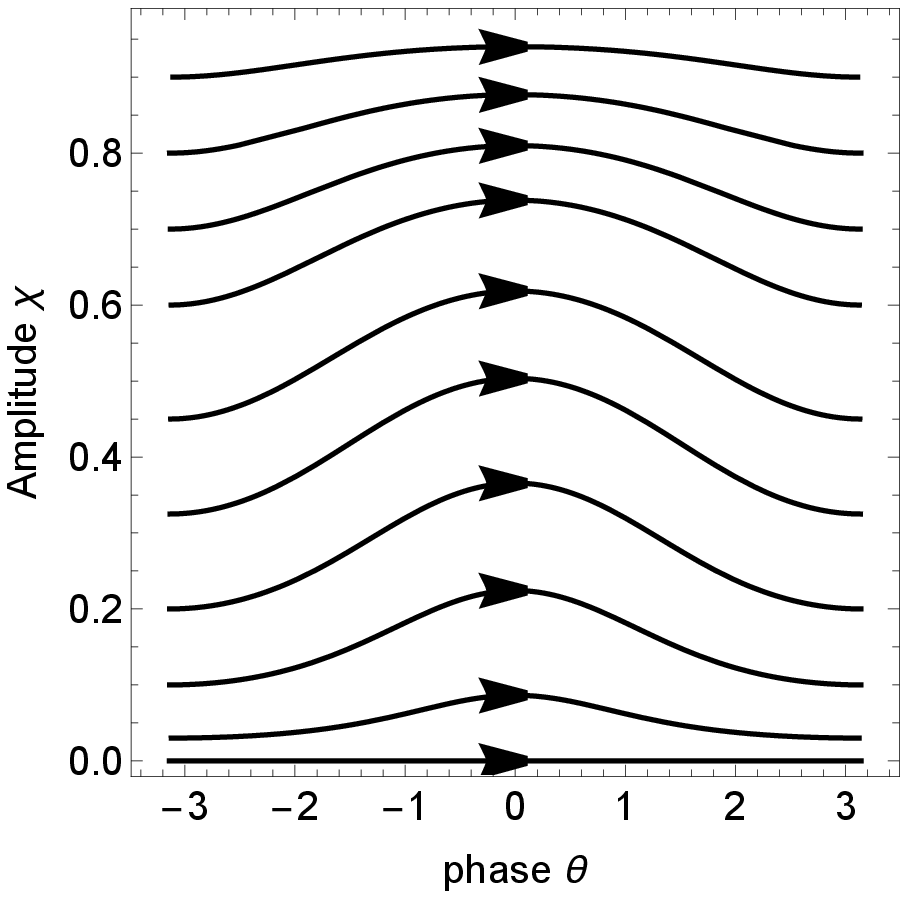}

\caption{\label{fig:PortraitPhaseTheo} (Color online) Phase portrait of the dynamical system~\eqref{eq:equaampldim3} in the plane $(\theta,\chi)$. Left plot : AP case, $\K = 0.2$. The separatrix ($J = 1/8 - \K$) is plotted in red, the black dotted line connects the points of vanishing $\dot\theta$. The red dots indicate the saddle points on the separatrix, the blue dot indicates the node~\eqref{eq:fpnode} which corresponds to $J = -{\K}\left(\K + 1\right)/2$. Closed trajectories (blue lines) are observed for $-{\K}\left(\K + 1\right)/2 < J < 1/8 - \K$. Open trajectories (black lines) are observed for $1/8 - \K < J < 1/8$. Right plot, NR case, $\K = 0.7$. All trajectories are open, $1/8 - \K < J < 1/8$.} 
\end{figure}

The phase portraits are shown in Fig.~\ref{fig:PortraitPhaseTheo}, in the left plot for the case of autoparametric resonance and in the right plot for the other case. This figure may be compared to Fig.~\ref{fig:portraitmathieu}~(b). As seen in the left plot of Fig.~\ref{fig:PortraitPhaseTheo}, the amplitude along the separatrix (solid red line) increases from $\chi = 0$ up to a maximum amplitude $\chi = 1-2\K$. Therefore, for any initial condition with very small amplitude, $0< \chi_0 \ll1$, the relevant phase space trajectory is very close to the separatrix so that $\chi$ eventually reaches a finite value very close to the maximum of the separatrix. Since $\chi$ is basically the amplitude of the  relative motion in the dimer, this evidences a parametric amplification of the relative motion induced by the center of mass motion, hence the phenomenon of autoparametric resonance. Along such a trajectory, the energy is given to the relative motion for $-\pi < \theta <0$, and restituted to the center of mass motion in the next half period, without any net energy transfer when averaged on the slow time scale. Formally, this corresponds to the fact that the amplitude equations~\eqref{eq:Amplitude} exhibits \emph{two} constant of the motions, whereas the underlying system~\eqref{eq:FuscoExact} has only \emph{one} constant of the motion, its conserved energy. In the same fashion, the amplitude equation~\eqref{eq:amplitudeparametric} for the parametrically forced Duffing oscillator has one constant of the motion, whereas the original system~\eqref{eq:pendulum} studied in  Sec.~\ref{sec:mathieu} is not conservative.

\subsubsection{Comparison with numerical simulations}
\label{sec:wellcompa}

In this section we compare our analytic results, based on the amplitude equation~\eqref{eq:Amplitude} to direct numerical simulations of the dynamical system~\eqref{eq:FuscoExact}. In the simulations, we take as initial conditions $x = 0$, $y = 0$ and non vanishing velocities. Our small parameter is thus
\beq
\epsilon^2 \equiv \left.\frac{m \left( {\dot x_1}^2 + {\dot x_2}^2\right)}{8 U_0} \right\vert_{(t = 0)}= \frac{{\dot x}_0^2 + {\dot y}_0^2}{4},
\label{eq:defesp}
\eeq
where the last expression is in dimensionless variables. To be consistent with the perturbative result~\eqref{eq:softorder1}, we have $x(t) = 2 \epsilon a \sin(t + \phi)$ and $y(t) = 2 \epsilon b \sin( t+ \psi)$, so that the initial condition implies vanishing phases $\phi$ and $\psi$ at $t = 0$. There remains no free parameter to undertake the comparison between the simulations and the multiple scale analysis. The determination of the slowly varying amplitudes and phases from the rough numerical data is explained in Appendix~\ref{App:Measuring}.

\begin{figure}[htb]
\center
\includegraphics[width=0.75\textwidth]{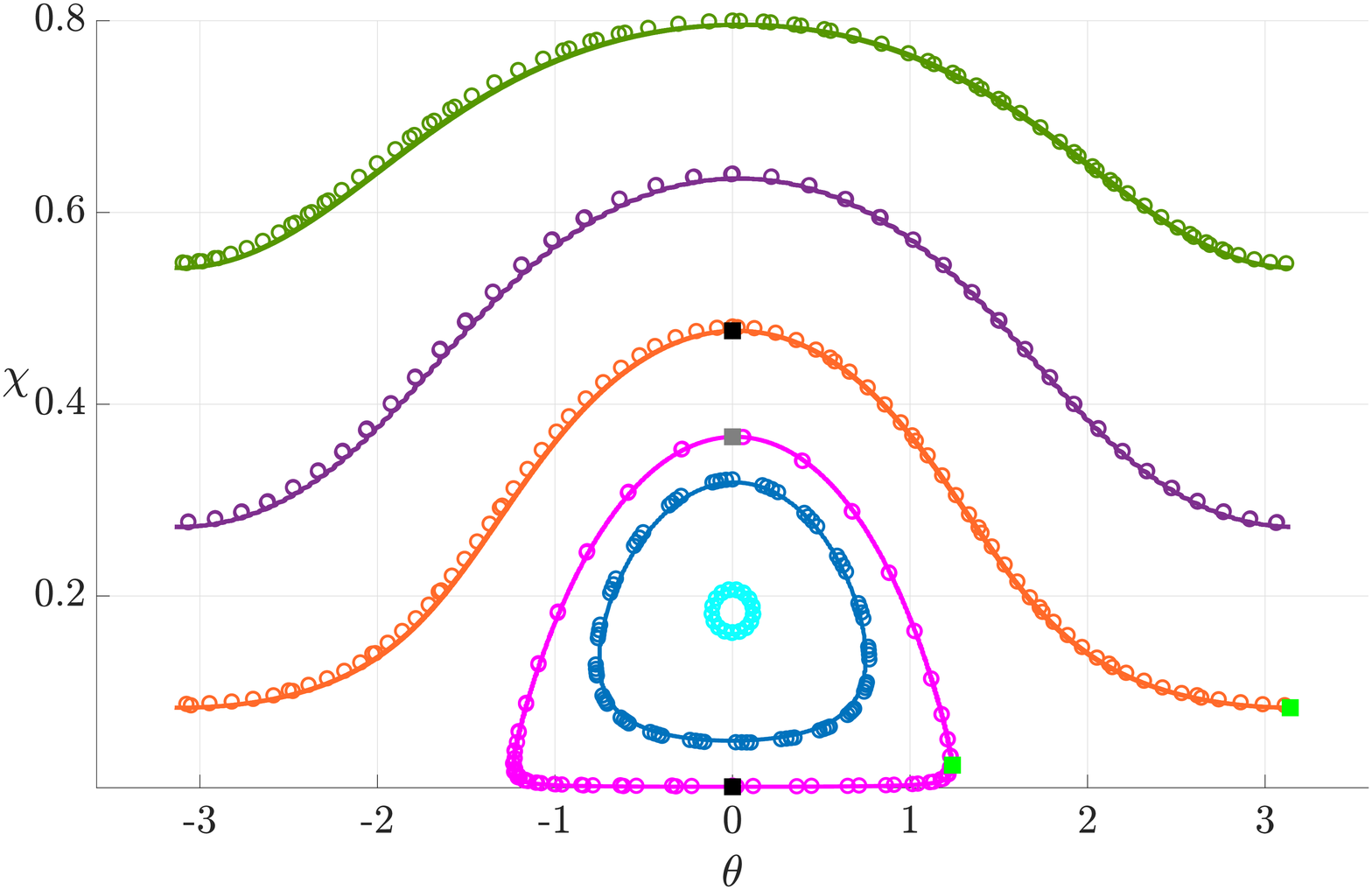}

\includegraphics[width=0.75\textwidth]{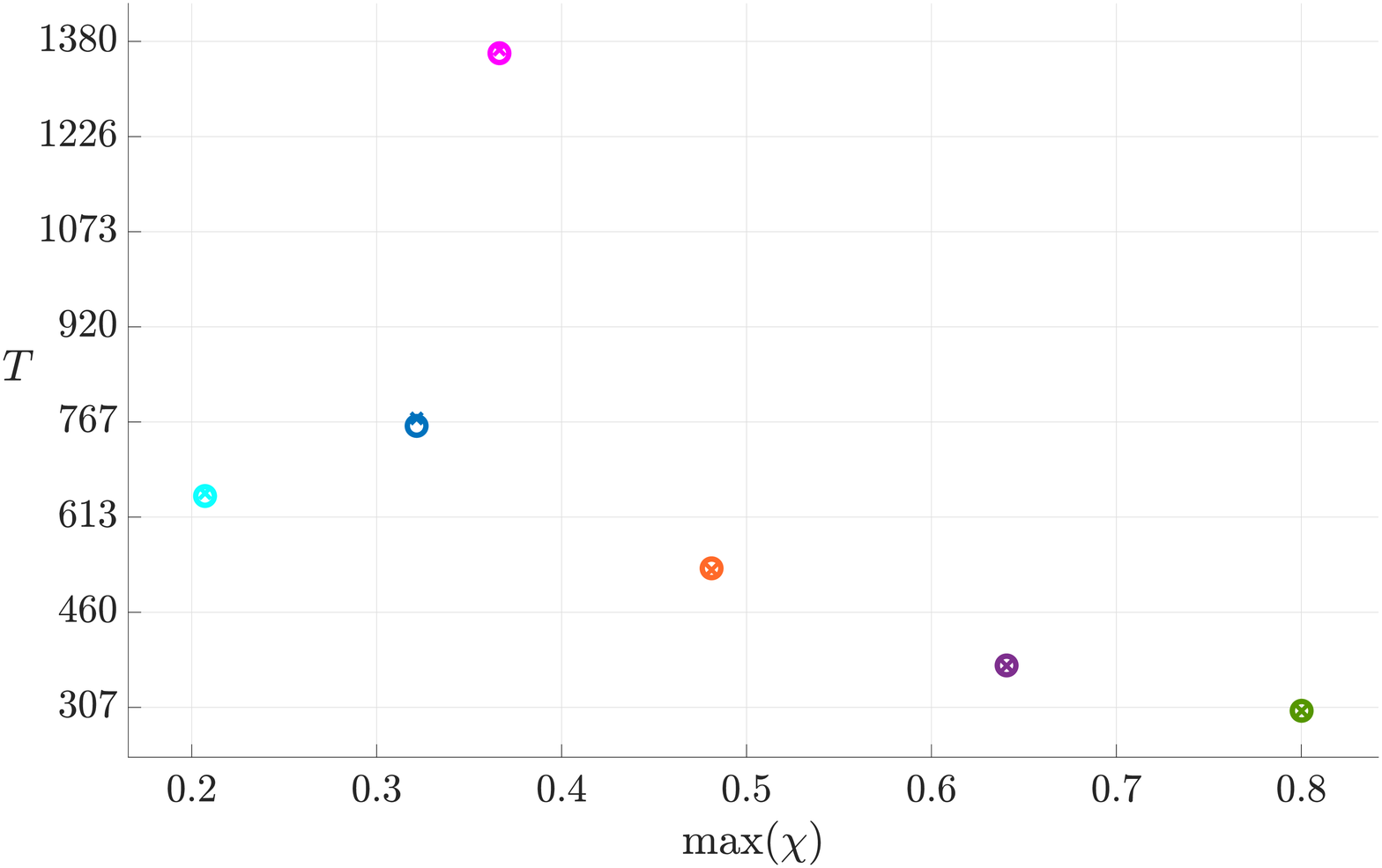}

\caption{\label{fig:PortraitPhaseCalc} (Color online). Top~: Phase trajectories in the plane $(\theta,\chi)$ obtained from Eqn.~\eqref{eq:solchiplusmoins} (dotted lines) and from direct numerical integration of~\eqref{eq:FuscoExact} (solid lines) for $\K = 0.158$ and $\epsilon = 0.224$. The initial phase is $\theta_0 = 0$ in all simulations. The initial values $\chi_0$ are~: $\chi_0 = 0.8$ ($J =0.013$),   $\chi_0 = 0.64$ ($J =-0.047$), $\chi_0 = 0.48$ ($J = - 0.082$), $\chi_0 = 0.32$ ($J = -0.091$),   $\chi_0 = 0.16$ ($J =-0.075$), $\chi_0 = 0.001$ ($J =-0.034$). The filled squares corresponds to the zooms displayed in Fig.~\ref{fig:evoltimedata}. Bottom~: Slow oscillations period $T$ (dimensionless) as a function of the maximum value of $\chi$ on the corresponding phase trajectory. Open circles are measured on direct numerical integration of~\eqref{eq:FuscoExact}, crosses are obtained from Eqn.~\eqref{eq:solchiplusmoins} (same color code as in the top plot).} 
\vskip -0.5cm
\end{figure}

In Fig.~\ref{fig:evoltimedata} we plot the relevant values of $x(t)$ and $y(t)$ for the normal modes as a function of time from a direct numerical integration of the system~\eqref{eq:FuscoExact} and compared them to a numerical integration of the amplitude equations~\eqref{eq:equaamplchi}. We see that the amplitude equations accurately predicts the slow modulation of the normal modes, without any fitting parameter. Moreover, we display zooms on both oscillatory modes in order to evidence the phase difference. Initially, the phase difference vanishes. For an open trajectory in phase space (upper plot of Fig.~\ref{fig:evoltimedata}, see Fig.~\ref{fig:PortraitPhaseCalc} for the relevant phase trajectory) when the amplitude of $x(t)$ is maximal the amplitude of $y(t)$ is minimal and the two signals are in quadrature ($\theta/2 = \pi/2$). For a closed trajectory (bottom plot of Fig.~\ref{fig:evoltimedata}, see Fig.~\ref{fig:PortraitPhaseCalc} for the relevant phase trajectory) the phase difference is clearly less than $\pi/2$. This illustrates the link between the phase difference $\theta$ and the time evolution of $x(t)$ and $y(t)$.

In the upper plot of Fig.~\ref{fig:PortraitPhaseCalc}, we compare the phase space trajectories calculated as explained in Appendix~\ref{App:Measuring} from direct numerical simulations of the system~\eqref{eq:FuscoExact} to the phase space trajectories provided by the multiple scale analysis, Eqn.~\eqref{eq:equaphase}. The relevant dimensionless stiffness is in the parametrically unstable tongue, $\K = 0.158$. Our data evidence a very good agreement with the multiple scale analysis. In the bottom plot of Fig.~\ref{fig:PortraitPhaseCalc}, we compare the slow modulation period measured from direct numerical simulations of the system~\eqref{eq:FuscoExact} (open circles) to a calculation derived from Eqn.~\eqref{eq:Horakxi} (crosses). Both calculations are in excellent agreement. The period is an increasing function of the maximum amplitude $\chi_{max}$ for closed trajectories and a decreasing function of $\chi_{max}$ for open trajectories, and it diverges on the separatrix.

\begin{figure}[htb]
\center
\includegraphics[width=0.8\textwidth]{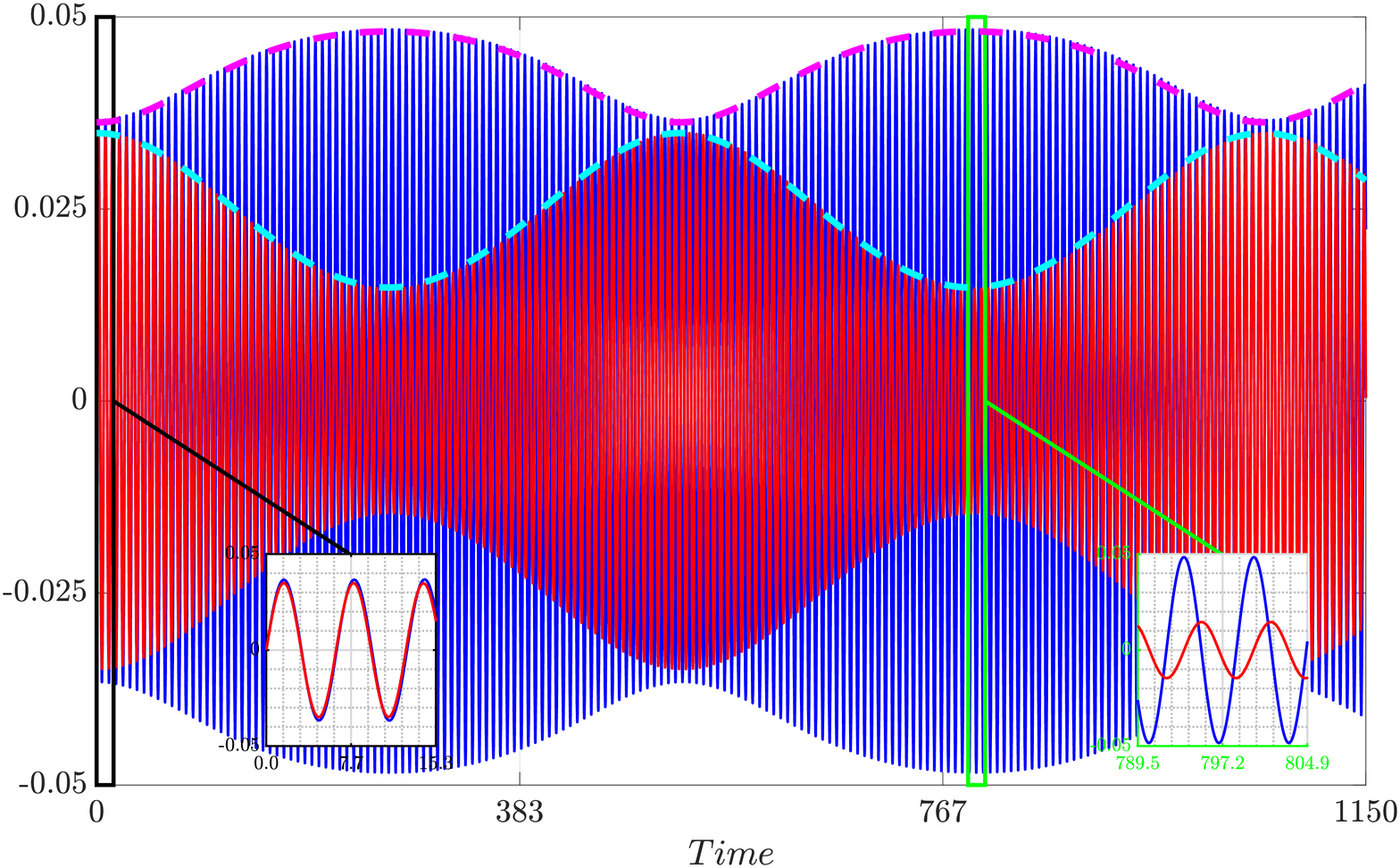}
\includegraphics[width=0.8\textwidth]{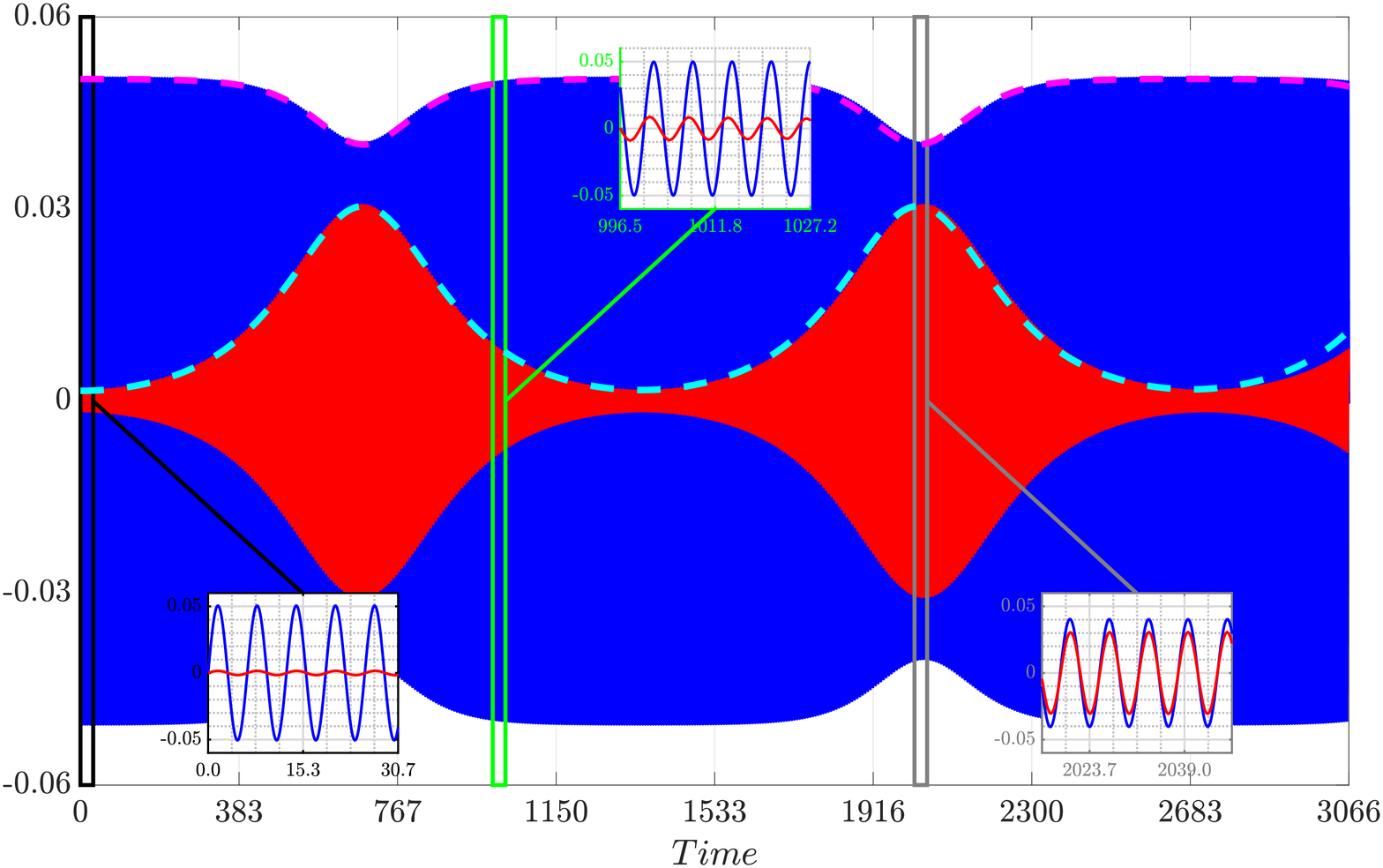}

\caption{\label{fig:evoltimedata} (Color online). Plot of $x(t)$ (dimensionless; solid blue lines) and $y(t)$ (dimensionless; solid red lines) as functions of the dimensionless time $t$, from a numerical integration of~\eqref{eq:FuscoExact} for $\K = 0.158$ and $\epsilon = 0.224$. The dashed lines are the slowly varying amplitudes of $x(t)$ (magenta) and $y(t)$ (cyan), from a numerical integration of the amplitude equations~\eqref{eq:equaamplchi}, with initial phase $\theta_0 = 0$. The insets are zooms that correspond to the points of the phase space trajectories indicated by filled squares in Fig.~\ref{fig:PortraitPhaseCalc}. Upper plot~: $\chi_0 = 0.52$ ($J = - 0.082$), open trajectory in Fig.~\ref{fig:PortraitPhaseCalc}.  The right hand zoom evidences that $x(t)$ and $y(t)$ are in quadrature, so that $\theta = \pi$. Bottom plot~: $\chi_0 = 0.999$ ($J = - 0.034$), closed trajectory in Fig.~\ref{fig:PortraitPhaseCalc}. The center zoom evidences that the phase difference between $x(t)$ and $y(t)$ is less than $\pi/2$. Note also the abcissae ranges.} 
\end{figure}

\clearpage

\subsection{Sliding dimer.}
\label{sec:sliding}

Let us now consider the case of the sliding dimer. We assume that the initial kinetic energy of the center of mass is much larger  than the depth of a potential well, so that the dimer motion is basically a monotonous translation of the center of mass together with oscillations of the particles around their equilibrium distance $a$. This amounts to assuming a strong enough interaction between the particles, in such a way that they cannot be distant from more than one period of the external potential. More general motions are studied numerically in a forthcoming paper~\cite{Maddi22papier2}.

Setting the same dimensionless variables as before, and expanding the interaction potential up to order four in the small quantity $y$, we get the dimensionless equations of motion
\beq
\begin{cases}  \ddot x = -  \sin x \cos y, \\ \ddot y = - K_2 y  + K_3 y^2 - K_4 y^3 -  \cos x \sin y.
\end{cases}
\label{eq:eqmvyanhaadim}
\eeq
where $K_2 = (U_{int}'' /U_0)(a/2 \pi)^2$,  $K_3 = (U_{int}^{(3)} /2U_0)(a/2 \pi)^3$ and  $K_4 = (U_{int}^{(4)} /6U_0)(a/2 \pi)^4$. Let us emphasize the contrast with the dimer in a well configuration. In this case, the autoparametric resonance happens for a soft bond, so that the harmonic approximation of the interaction potential is sufficient. For a sliding dimer, the autoparametric resonance  requires a strong bound, and consistently nonlinear terms in the interaction potential expansion are to be taken into account. Note also that we have taken into account the fact that any reasonable interaction potential such as the Lennard-Jones potential~\cite{FuscoEPJB03}, or such as the Morse potential which rather well describes chemical bonds~\cite{LeRoy07}  is dissymmetric near its minimum which requires the cubic term proportional to $K_3$. 

Let $V_0$ be the initial velocity of the center of mass, in dimensionless units. Introducing $\tau = V_0 t$, the previous equation now reads
\beq
\begin{cases}  \ddot x = - \dfrac{ 1}{ V_0^2} \sin x \cos y, \\[-15pt] \\ \ddot y = - \dfrac{K_2}{V_0^2} y + \dfrac{K_3}{V_0^2} y^2 - \dfrac{K_4}{V_0^2} y^3 - \dfrac{ 1}{ V_0^2} \cos x \sin y,
\end{cases}
\label{eq:eqmvyanhaadim}
\eeq
where from now on $\dot x \equiv \dd x/\dd \tau$.

In the sliding configuration,  the kinetic energy of the center of mass is much larger  than the depth of the potential, $8 U_0/(m \hat{V}_0^2) \ll 1$ (where $\hat{V}_0$ is the velocity in SI units) which in dimensionless units means $1/ V_0^2 \ll 1$. Since it is the  center of mass motion $x(\tau)$ that sets the velocity, it cannot be considered as small. In contrast, we assume a small amplitude of the relative motion, $\vert y \vert \ll 1$. This is obviously not the most general configuration, but we may expect to obtain a perturbative description of such motions. To sum-up, we introduce the following expansions,
\beq
\begin{cases} x(\tau) = X_0 + \epsilon X_1 + \epsilon^2 X_2 + \ldots, \\
y(\tau) = \epsilon Y_1 + \epsilon^2 Y_2 + \epsilon^3 Y_3 + \ldots, \end{cases}
 \label{eq:echellesXY}
 \eeq
 \beq
\frac{\dd }{\dd \tau} = \frac{\partial}{\partial T_0} + \epsilon \frac{\partial}{\partial T_1} + \epsilon^2 \frac{\partial}{\partial T_2} \donc \frac{\dd^2 }{\dd \tau^2} = \frac{\partial^2}{\partial T_0^2} + 2 \epsilon \frac{\partial^2}{\partial T_0\partial T_1} + \epsilon^2 \left(2\frac{\partial^2}{\partial T_0\partial T_2} + \frac{\partial^2}{\partial T_1^2}\right) ,
\label{eq:echellesT}
\eeq
where $\epsilon^2  \equiv  1/V_0^2 ={U_0}/({m \hat{V}_0^2}) \ll 1$. We describe the stiff spring by setting
\beq
\dfrac{K_2}{V_0^2} =  \frac{1}{4} - \eta \epsilon^2, \qquad \dfrac{K_3}{V_0^2} =  \kappa_3, \qquad \dfrac{K_4}{V_0^2} =  \kappa_4,
\label{eq:devpara}
\eeq
where the parameters $\eta$, $\kappa_3$ and $\kappa_4$ are assumed to be of order ${\cal O}(\epsilon^0)$. The chosen value of $K_2$ is convenient to describe the main parametric resonance.

Taking into account all relevant orders, the system~\eqref{eq:eqmvyanhaadim} is writen perturbatively as
\beqa
&\quad&\left[\frac{\partial^2}{\partial T_0^2} + 2 \epsilon \frac{\partial^2}{\partial T_0\partial T_1} + \epsilon^2 \left(2\frac{\partial^2}{\partial T_0\partial T_2} + \frac{\partial^2}{\partial T_1^2}\right)\right](X_0 + \epsilon X_1 + \epsilon^2 X_2) = \nonumber \\ &\quad&\qquad\qquad\qquad\qquad =  -\epsilon^2 \sin(X_0 + \epsilon X_1 + \epsilon^2 X_2)\cos\left(\epsilon Y_1 + \epsilon^2 Y_2 + \epsilon^3 Y_3\right),
\label{eq:eqmvyanhaadimX}
\eeqa
and
\beqa
&\;&\left[\frac{\partial^2}{\partial T_0^2} + 2 \epsilon \frac{\partial^2}{\partial T_0\partial T_1} + \epsilon^2 \left(2\frac{\partial^2}{\partial T_0\partial T_2} + \frac{\partial^2}{\partial T_1^2}\right)\right](\epsilon Y_1 + \epsilon^2 Y_2 + \epsilon^3 Y_3) =   \nonumber \\ &\quad&\quad = -\left(\frac{1}{4} - \eta \epsilon^2\right)(\epsilon Y_1 + \epsilon^2 Y_2 + \epsilon^3 Y_3) + \kappa_3 (\epsilon Y_1 + \epsilon^2 Y_2 + \epsilon^3 Y_3)^2 -  \nonumber \\ &\quad&\qquad - \kappa_4 (\epsilon Y_1 + \epsilon^2 Y_2 + \epsilon^3 Y_3)^3 - \epsilon^2 \cos(X_0 + \epsilon X_1 + \epsilon^2 X_2 )\sin\left(\epsilon Y_1 + \epsilon^2 Y_2  + \epsilon^3 Y_3\right).
\label{eq:eqmvyanhaadimY}
\eeqa

In this section, we take advantage of the versatility of the multiple scale method, since the motion of the center of mass is not oscillatory but basically a translation. The calculations are thus explained in more details than in the previous sections.

\subsubsection*{Order ${\cal O}(\epsilon^0)$.}

At this order, the only contribution come from Eqn.~\eqref{eq:eqmvyanhaadimX}, and is readily solved
\beq
\frac{\partial^2 X_0}{\partial T_0^2} = 0, \quad \frac{\partial X_0}{\partial T_0} = A_0(T_1,\ldots),\quad X_0 = A_0(T_1,\ldots)T_0.
\label{eq:slidingYsolu0}
\eeq
The initial conditions will be used at the end of the calculation, and for now $A_0(T_1,\ldots)$ is an unknown function of the slow scales $T_1,T_2,\ldots$. Nevertheless, it is important to keep in mind that, because of the definition of the time scale $\tau$, $A_0 = 1+ \ldots$, where the dots stand for a small correction that will be found later to be of order $\epsilon^2$.

\subsubsection*{order ${\cal O}(\epsilon)$.}

The relevant terms from Eqn.~\eqref{eq:eqmvyanhaadimX} reads
\beq
\frac{\partial^2 X_1}{\partial T_0^2} = - 2 \frac{\partial^2 X_0}{\partial T_0 \partial T_1} = -2 \frac{\partial A_0}{\partial T_1}.
\label{eq:ordre1X}
\eeq
To get a consistent expansion, the term $X_1$ should not increase faster with $T_0$ than $X_0$, which requires  ${\partial A_0}/{ \partial T_1} = 0$ so that $A_0(T_2,\ldots)$. Then we take without loss of generality $X_1 = 0$ since the relevant initial condition on $x(t)$ may be set on $X_0$.

The term of order ${\cal O}(\epsilon)$ that comes from~\eqref{eq:eqmvyanhaadimY},  reads
\beq
\frac{\partial^2 Y_1}{\partial T_0^2} + \frac{1}{4}Y_1 = 0, \quad Y_1 = B(T_1)e^{i T_0/2} + \overline{B}(T_1)e^{-i T_0/2}.
\label{eq:dimereO1}
\eeq

\subsubsection*{Order ${\cal O}(\epsilon^2)$.}

At this order, we get the second order correction for $x$, which reads
\beq
\frac{\partial^2 X_2}{\partial T_0^2} = -\sin X_0 - 2 \frac{\partial^2 X_0}{\partial T_0 \partial T_2} = -\sin (A_0 T_0) - 2\frac{\partial A_0}{ \partial T_2}.
\label{eq:dimereO2}
\eeq
As before, to get a consistent expansion, the term $X_2$ should not increase faster with $T_0$ than $X_0$, which requires ${\partial A_0}/{ \partial T_2} = 0$. Therefore, the solution at this order reads
\beq
X_0 = A_0(T_3)T_0, \quad X_2 = \frac{1}{A_0(T_3)^2} \sin[A_0(T_3)T_0].
\label{eq:dimereO2solu}
\eeq
The second order correction $X_2$ is thus independent on $T_2$ and $T_1$, which will be used later.

The term of order ${\cal O}(\epsilon^2)$ that comes from~\eqref{eq:eqmvyanhaadimY},  reads
\beq
\frac{\partial^2 Y_2}{\partial T_0^2} + \frac{1}{4}Y_2 = -2 \frac{\partial^2 Y_1}{\partial T_0 \partial T_1} +\kappa_3 Y_1^2 = - i \frac{\partial B}{\partial T_1}e^{i T_0/2} + \kappa_3\left(B^2 e^{i T_0} +  \vert B \vert^2 \right) + CC,
\label{eq:dimereO2Y}
\eeq
where "$CC$" means "complex conjugate". The resonant term must vanish, therefore $\partial B/\partial T_1 = 0$, and we get at this order
\beq
Y_2 = 8 \kappa_3 \vert B(T_2)\vert^2 - \frac{4 \kappa_3}{3}B(T_2)^2 e^{i T_0}  - \frac{4 \kappa_3}{3}\overline{B}(T_2)^2 e^{-i T_0} .
\label{eq:dimereO2Ysolu}
\eeq

\subsubsection*{order ${\cal O}(\epsilon^3)$.}

These terms occur in the relative motion equation~\eqref{eq:eqmvyanhaadimY}, which reads
\beq
\frac{\partial^2 Y_3}{\partial T_0^2} + \frac{1}{4}Y_3 = - 2 \frac{\partial^2 Y_1}{\partial T_0 \partial T_2} + \eta Y_1 + 2 \kappa_3 Y_1 Y_2 - \kappa_4 Y_1^3 - Y_1 \cos X_0
\label{eq:dimereO3brut}
\eeq
To obtain the required amplitude equation for the relative motion, we only need the calculation of the secular term in the right hand member. It reads
\beq
\frac{\partial^2 Y_3}{\partial T_0^2} + \frac{1}{4}Y_3 = e^{i T_0/2}\left[- i \frac{\partial B}{\partial T_2} + \eta B - \left(3\kappa_4  - \frac{40 \kappa_3^2}{3}\right) B^2 \overline{B} - \frac{\overline{B}}{2}e^{i(A_0 - 1)T_0}\right] + CC + NST
\label{eq:dimereO3}
\eeq
where $NST$ means \emph{non secular terms}. The last secular term comes from the coupling between the center of mass and relative motions. It is emphasized as the boxed term in
\beqa
2Y_1 \cos X_0 &=& \left(B e^{i T_0/2} + \overline{B} e^{-i T_0/2}\right)\left(e^{i A_0T_0} + e^{-i A_0 T_0}\right) = \nonumber \\ &=& B e^{i \left(A_0 + \frac{1}{2}\right)T_0} + \boxed{\overline{B}e^{i \left(A_0 - \frac{1}{2}\right)T_0}} + B e^{- i \left(A_0 - \frac{1}{2}\right)T_0} + \overline{B} e^{-i \left(A_0 + \frac{1}{2}\right)T_0}. \nonumber
\eeqa
The term in the box is indeed a secular term, because of the initial condition on the center of mass motion.  Indeed, to be consistent with the center of mass motion at order $\epsilon^2$, we have
$$
x(\tau)  = A_0 \tau + \frac{\epsilon^2}{A_0^2}\sin A_0 \tau \donc \frac{\dd x}{\dd \tau} = A_0 + \frac{\epsilon^2}{A_0}\cos A_0 \tau.
$$
In the variable $\tau$, the initial velocity is unity, so that
$$
\left.\frac{\dd x}{\dd \tau}\right\vert_{\tau = 0} = A_0 + \frac{\epsilon^2}{A_0} = 1 \donc A_0 = 1 - \epsilon^2 + {\cal O}(\epsilon^4).
$$
Injecting this expression for $A_0$ in the secular term of Eqn.~\eqref{eq:dimereO3}, we get the relevant amplitude equation for the relative motions  as the solvability condition
\beq
\frac{\partial B}{\partial T_2} =- i \eta B + i \kappa B^2 \overline{B} + i \frac{\overline{B}}{2}e^{- i  T_2}, \quad \kappa \equiv 3 \kappa_4  - \frac{40 \kappa_3^2}{3}.
\label{eq:formenormaleB}
\eeq
Setting $B = e^{- i  T_2/2}D$, we recover the autonomous amplitude equation for the parametric instability,
\beq
\frac{\partial D}{\partial T_2} = i \left(\frac{1}{2} - \eta \right) D +  i \kappa \vert D \vert^2 D  + i \frac{\overline{D}}{2}.
\label{eq:formenormaleD}
\eeq
which is the same as Eqn.~\eqref{eq:amplitudeparametric}, apart from small changes in the notations. Note that the sign of $\kappa$ is not relevant, since we can take the complex conjugate of this equation as well~\cite{Fauve94}.

Let us introduce a real amplitude $R$ and phase $\phi$, setting $D \equiv R e^{i\phi}$. The constant of the motion $\widetilde{H}$
\beq
\widetilde{H} =  -  R^2\left(\frac{1}{2}\kappa R^2 +  \frac{1}{2}\cos 2\phi +  \frac{1}{2}- \eta\right),
\label{eq:hamiltonianparametricmod}
\eeq
takes the place of $H$ in~\eqref{eq:hamiltonianparametric}. The condition of parametric amplification of the relative motion now reads
\beq
0 \leqslant \eta  \leqslant 1.
\label{eq:resparadim}
\eeq

\begin{figure}[htb]
\center
\includegraphics[width=0.33\textwidth]{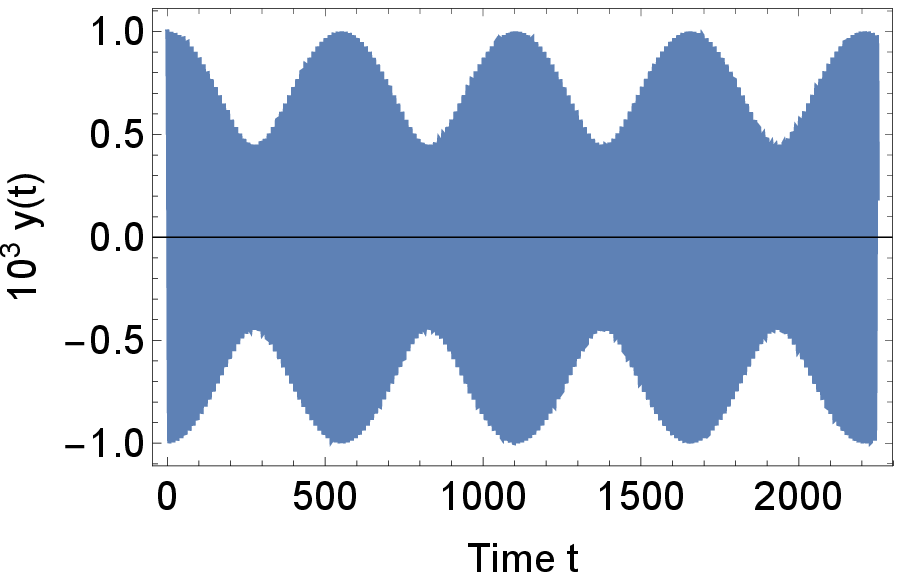}\includegraphics[width=0.33\textwidth]{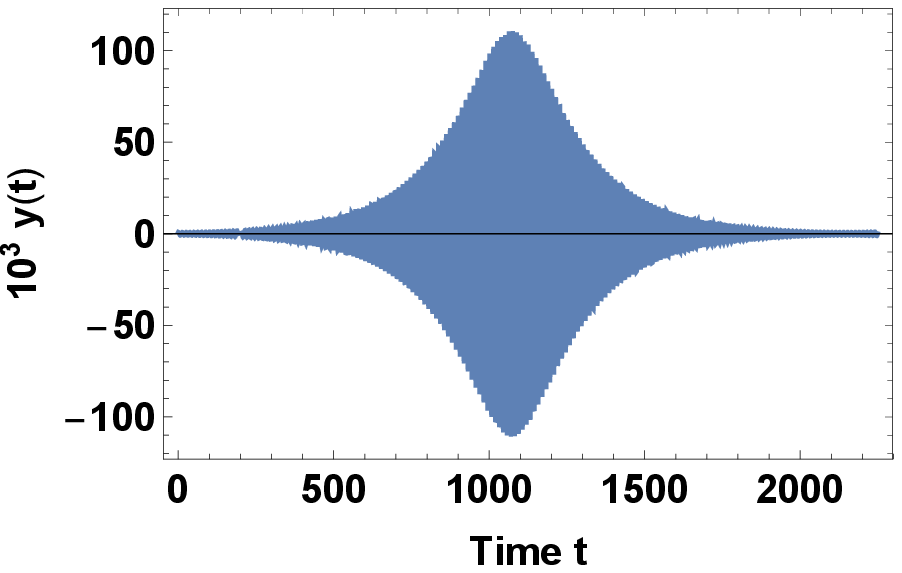}\includegraphics[width=0.33\textwidth]{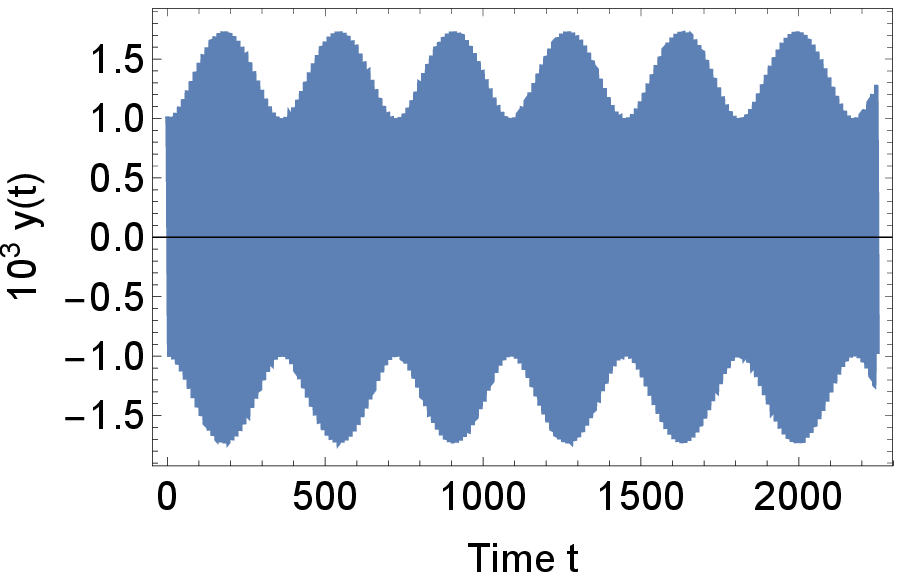}

\caption{\label{fig:DimereSlidingRP} (Color online). Plot of the dimensionless amplitude $10^3 y(t)$ of the relative motion as a function of the dimensionless time $t$ from a numerical integration of the system~\eqref{eq:eqmvyanhaadim}. The relevant parameters are $\kappa = 1$ and  $\epsilon = 0.1$. The initial conditions for the numerical integration  are $x(0) = 0$, $\dot x(0) = 1$, $y(0) = 0.001$ and $\dot y(0) = 0$. From left to right the spring constant  $1/4 - \epsilon^2 \eta$ is $0.2375$ ($\eta = 1.25$), $0.245$ ($\eta = 0.5$) and $0.255$ ($\eta = -0.5$). As expected, the relative motion inside  the parametric resonance tongue~\eqref{eq:resparadim} (center plot) is markedly different from the others which are outside the resonance tongue.} 
\end{figure}

The sliding motion of the center of mass may induce a parametric amplification of the particles relative motion. This parametric amplification in the sliding regime only happens when the interaction is strong enough for the constants $\kappa_i$ to be of the same order as the dimensionless kinetic energy $V_0^2$, as explained in~\eqref{eq:devpara}. The condition for parametric excitation of the relative motion by the center of mass motion is given by~\eqref{eq:resparadim}. We show the relative motion as a function of time for several values of $\eta$ in Fig.~\ref{fig:DimereSlidingRP}. The  width of the resonance tongue in the vicinity of ${K}_2/{V}_0^2 = 1/4$ does scale as $\epsilon^2$. In the center plot of Fig.~\ref{fig:DimereSlidingRP} the relevant value of $\eta = 1/2$ is inside the parametric resonance tongue, and the relative motion $y(t)$ as a function of time exhibits the expected parametric resonance, with an amplitude amplification by a factor 100. In the left and right plots of Fig.~\ref{fig:DimereSlidingRP} the values of $\eta$ are outside the parametric resonance tongue, and consistently  the relative motion $y(t)$ as a function of time exhibits no parametric resonance.

\begin{figure}[htb]
\center
\includegraphics[width=0.5\textwidth]{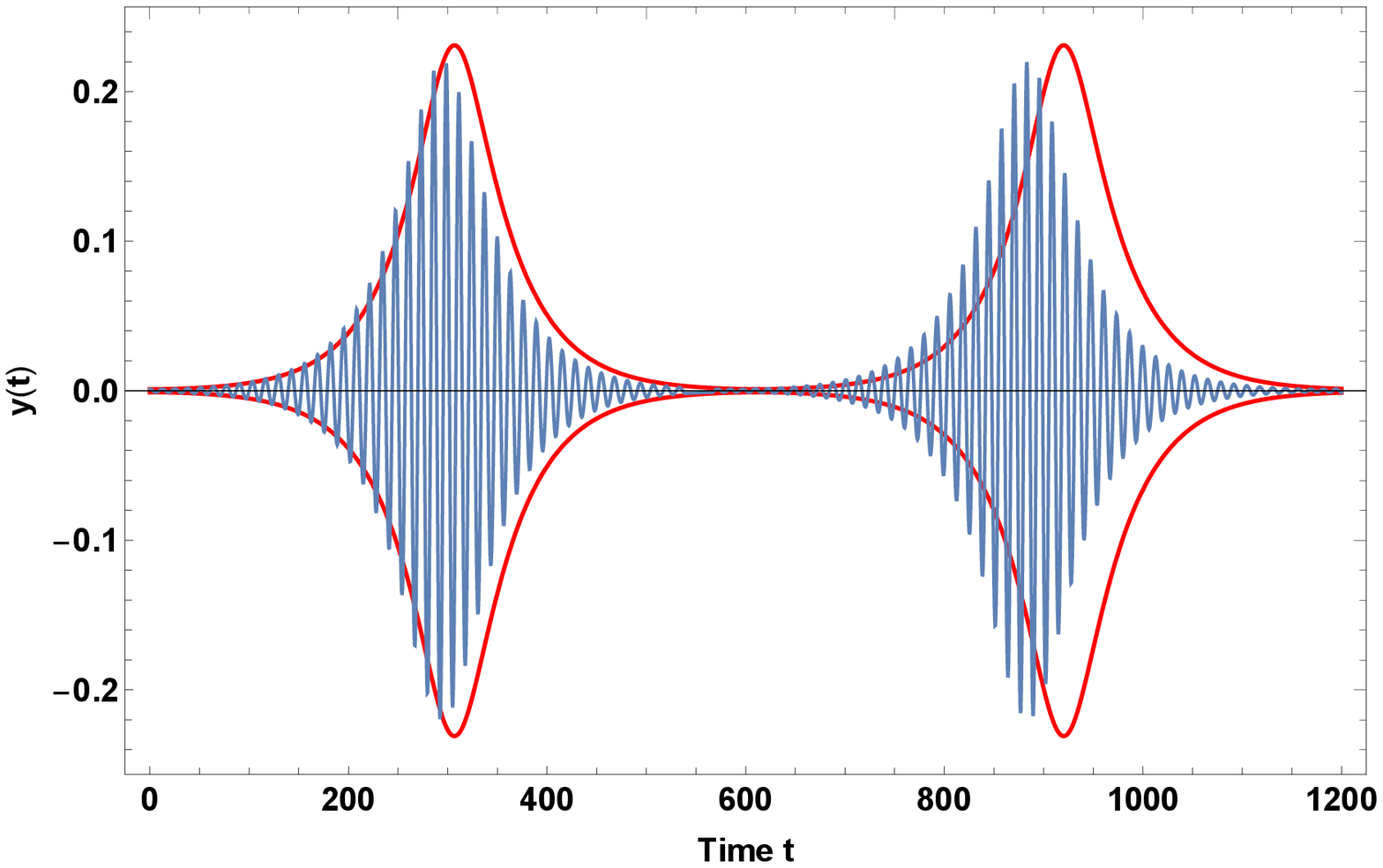}

\includegraphics[width=0.5\textwidth]{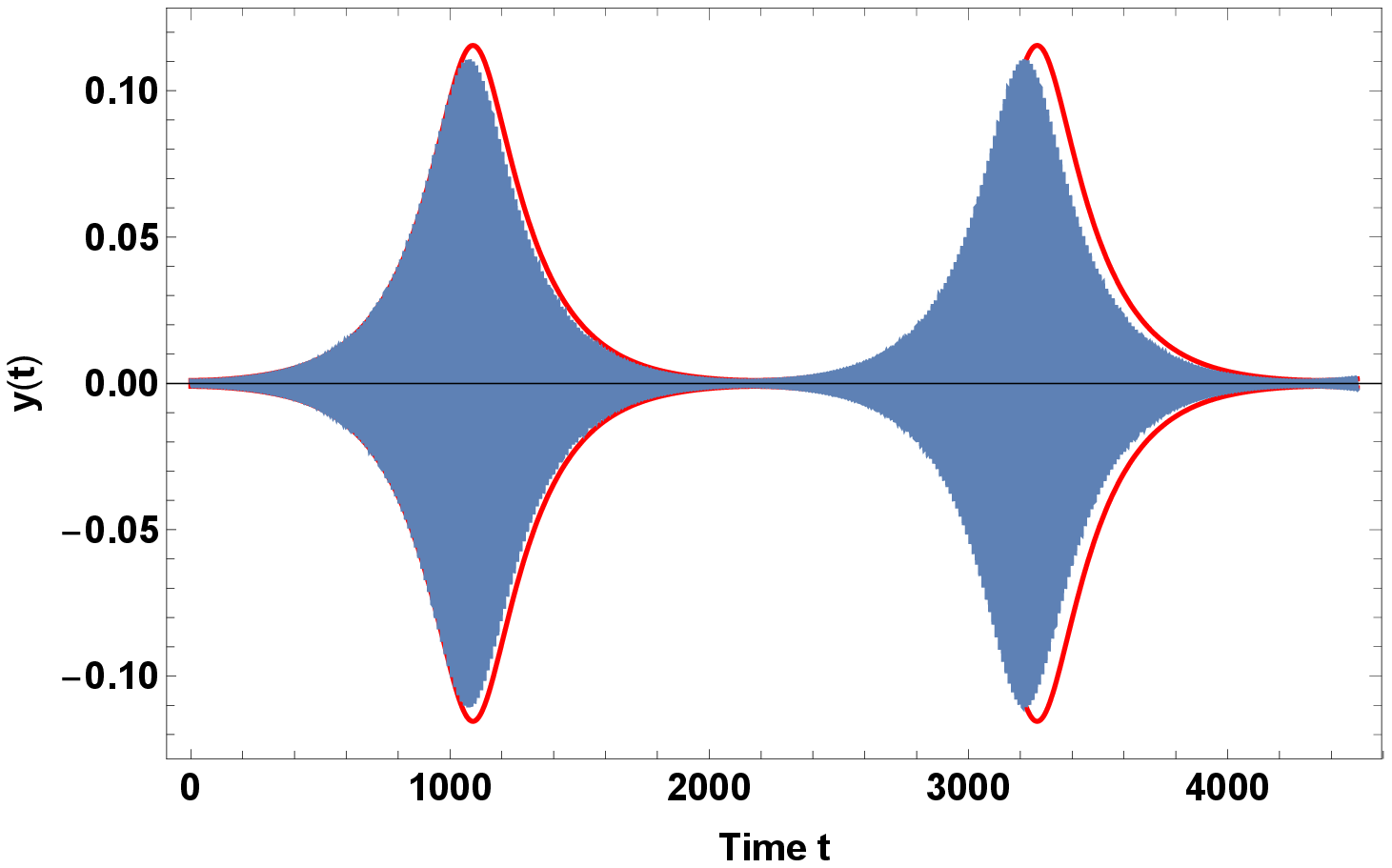}

\includegraphics[width=0.5\textwidth]{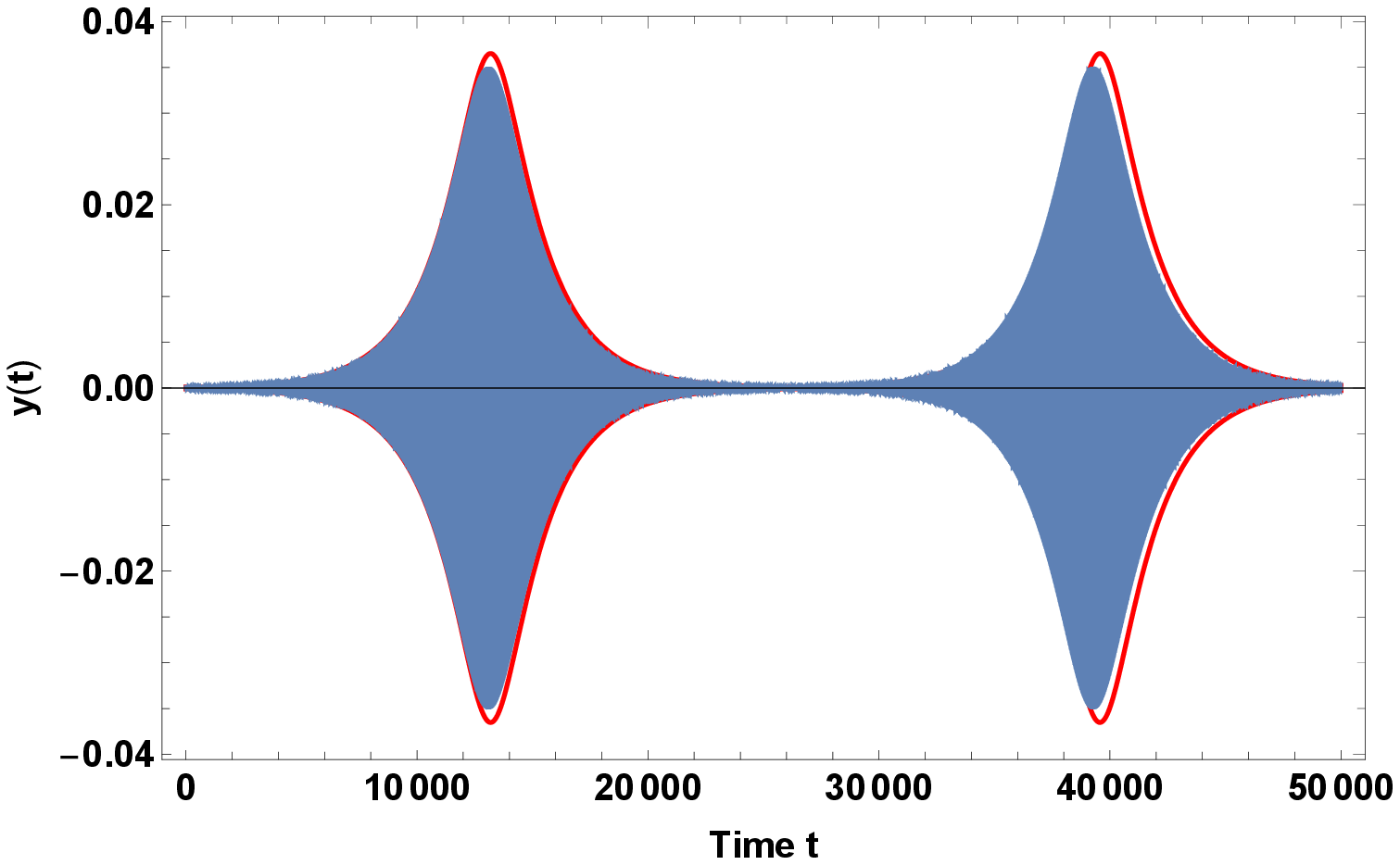}

\caption{\label{fig:DimereSliding} (Color online). Blue (dark grey) solid line~: Plot of the dimensionless amplitude $y(t)$ of the relative motion as a function of the dimensionless time $t$ from a numerical integration of the system~\eqref{eq:eqmvyanhaadim}. Red (light grey) solid line~: Amplitude $R(t) = \vert B(t)\vert$ of the quick oscillations, from a numerical integration of the amplitude equation~\eqref{eq:formenormaleD}. The relevant parameters are $\kappa = 1$ and $\eta = 1/2$, and from top to bottom $\epsilon = 0.2$,  $\epsilon = 0.1$ and $\epsilon = 0.032$. The initial conditions for the numerical integration of~\eqref{eq:eqmvyanhaadim} are $x(0) = 0$, $\dot x(0) = 1$, $\dot y(0) = 0$, and for the numerical integration of~\eqref{eq:formenormaleD} we set $\phi(0) = Arg(B)\vert_{t = 0} = 0$ and $R(0) = y(0)/(2 \epsilon)$. Note that the choice $\dot x(0) = 1$ ensures that the time unit is the same in all plots. From top to bottom the initial condition for $y$ is $y(0) = 0.001$, $y(0) = 0.001$ (same as before) and $y(0) = 0.0001$. In the last two plots the resolution is insufficient to distinguish the quick oscillations.} 
\end{figure}

In Fig.~\ref{fig:DimereSliding}, we compare the exact solution, given by the numerical integration of~\eqref{eq:eqmvyanhaadim} to the predictions of  the amplitude equation~\eqref{eq:formenormaleD}, for decreasing values of the small parameter $\epsilon$ (from top to bottom). We see that the time evolution of the relative motion amplitude is indeed very well predicted by the asymptotic analysis, and the smaller $\epsilon$ the better. The maximum amplitude of the relative motion scales as $\epsilon$, and the characteristic time for the parametric amplification of the relative motion scales as $\epsilon^2$ (see the time scales in the plots). The amplitude growth begins as an exponential, which evidences the parametric amplification. The ratio between the maximum amplitude $y_M$ of the relative motion and its initial value $y(0)$ is 200, 100 and 350 from top to bottom, which evidences a huge amplification of the relative motion. Since the system~\eqref{eq:eqmvyanhaadim} is conservative, the energy transfer between the sliding motion of the center of mass and the relative motion is periodic.


\section{Conclusion}
\label{sec:conclusion}

A dimer in a periodic potential is a simple system with a complicated dynamics. It is conservative, but not integrable. Its motions are determined by its initial energy and the stiffness of the interaction  between the particles. If the equilibrium length of the dimer is equal to the period of the potential, this commensurate configuration make detailed calculations achievable in two limits. 

In the first one, the initial energy and the interaction energy are small enough in comparison with the external potential energy barrier so that the dimer is trapped in a potential well. In this configuration, the center of mass motion may induce a parametric amplification of the relative motion. The system comes down to coupled nonlinear oscillators  that are easily addressed by a consistent multiple scale expansion. Moreover, the amplitude equations obtained with this analysis are found to be integrable, which allows a complete description of the dimer motions. When numerical simulations of the actual system are compared to the analytic description, this latter is found to describe accurately the motions of the dimer. It will be shown in a forthcoming paper~\cite{Maddi22papier2} that the validity of our amplitude equations extends on much higher values of the small parameter (the ratio between the initial energy and the energy barrier) than expected. This system therefore exhibits autoparametric resonance between two oscillatory modes of a conservative system on a rather large parameter range.

The second configuration allowing a complete analytical description is when both the initial kinetic energy and the interaction energy are high enough in comparison with the external potential energy barrier for the dimer to slide along the external potential. This assumes a strong bond between the particles in the dimer, but otherwise the interaction potential is treated in full generality. The basis motion is the sliding of the dimer center of mass, that is coupled by the external potential to the relative motions of the particles. Since the autoparametric resonance requires a strong bond, we expand the interaction potential up to the fourth order. Taking advantage of the versatility of the multiple scale expansion, we show that the relevant amplitude equation is exactly that of the parametrically forced Duffing equation, which is a paradigm of parametric amplification of a nonlinear oscillator. Since no external energy is provided to the system, this is another example of autoparametric behavior. 

Apart from these two limiting cases, the system may exhibit complicated behaviors for which the commensurability is lost, when the initial energy is high enough, and the interaction energy small enough for the particles to jump in non neighbouring potential wells. A description of such behaviors will be the subject of a forthcoming work~\cite{Maddi22papier2}. Another extension of this work is to take into account a dissipative term, together with a non zero temperature.

\clearpage

\appendix

\section{Constants of motion for a trapped dimer}
\label{App:Noether}

The equations~\eqref{eq:Amplitude} are Lagrange equations for the lagrangian
\beq
\lag = i\left(A \dot{\overline{A}} - \overline{A}\dot{A}\right) + i\left(B \dot{\overline{B}} - \overline{B}\dot{B} \right) + \frac{\vert A\vert^4 + \vert B\vert^4}{4}  + \vert A \vert^2 \vert B \vert^2 - 2 \K \vert B\vert^2 +\frac{B^2 \overline{A}^2 + A^2\overline{B}^2}{4},
\label{eq:lagrangien}
\eeq
This lagrangian~\eqref{eq:lagrangien} is obviously invariant under the transform
$$
A \longrightarrow A' = A e^{i \eta}, \quad \overline{A} \longrightarrow \overline{A}' = \overline{A} e^{-i \eta}, \quad B \longrightarrow B' = B e^{i \eta}, \quad \overline{B} \longrightarrow \overline{B}' = \overline{B} e^{-i \eta},
$$
where $\eta$ is a real constant phase. Assuming $\vert \eta \vert \ll 1$,  we get the relevant infinitesimal transform, so that we deduce from Noether's theorem~\cite{Goldstein80} the conserved quantity~\eqref{eq:norme},
$$
\frac{\partial \lag}{\partial \dot{{A}}} (i {A}) + \frac{\partial \lag}{\partial \dot{\overline{A}}} (- i\overline{A}) + \frac{\partial \lag}{\partial \dot{{B}}} (i {B}) + \frac{\partial \lag}{\partial \dot{\overline{B}}} (- i\overline{B}) = 2 \vert A \vert^2 + 2 \vert B \vert^2.
$$

Another conserved quantity is due to the fact that the lagrangian~\eqref{eq:lagrangien} has no explicit time dependence. The conserved quantity $H$ reads
$$
-H \equiv \dot{A} \frac{\partial \lag}{\partial \dot{A}} +\dot{\overline{A}}  \frac{\partial \lag}{\partial \dot{\overline{A}}} + \dot{B} \frac{\partial \lag}{\partial \dot{B}} +\dot{\overline{B}}  \frac{\partial \lag}{\partial \dot{\overline{B}}} - \lag,
$$
so that
\beq
H = 2 \K \vert B \vert^2 - \vert A \vert^2 \vert B \vert^2 - \frac{\vert A \vert^4 +\vert B \vert^4}{4} - \frac{1}{4}\left(B^2 \overline{A}^2 + A^2 \overline{B}^2\right)
\label{eq:constW}
\eeq
There must be a relationship between $H$, $N$ and the constant $J$, since only two of them may be independent. Indeed
$$
H = 2 \K (N - a^2) + \frac{a^4 + b^4}{4} - \underbrace{\left[a^2 b^2 + \frac{a^4 + b^4}{2} \right]}_{ = N^2/2} - \frac{1}{2} a^2 b^2 \cos \theta,
$$
so that eventually
$$
H - 2 \K N + \frac{N^2}{2} = 2\left(\frac{a^4 + b^4}{8} - \K a^2 - \frac{1}{4} a^2 b^2 \cos \theta\right) = 2 J.
$$

\section{About the constant $N$.}
\label{App:constanteN}

In all generality, the dimer motion depends on the initial conditions $x_0$, $y_0$, $\dot{x}_0$ et $\dot{y}_0$, in dimensional variables. For the analysis of Sec.~\ref{sec:well} to be relevant, the initial conditions must be such that
\beq
\frac{2 \pi \vert x_0\vert}{a} \ll 1, \quad \frac{2 \pi \vert y_0\vert }{a} \ll 1, \quad \sqrt{\frac{m}{U_0}} \vert\dot{x}_0 \vert\ll 1, \quad \sqrt{\frac{m}{U_0}} \vert\dot{y}_0\vert \ll 1.
\label{eq:inegdim}
\eeq

Let us keep the tilde $\widetilde{\;}$ for the dimensionless variables, for the sake of clarity. Let us define
\beq
\xt \equiv \frac{2 \pi  x_0}{a},\quad \yt \equiv \frac{2 \pi  y_0}{a},\quad \dxt \equiv \sqrt{\frac{m}{U_0}} \dot{x}_0,\quad \dyt \equiv \sqrt{\frac{m}{U_0}} \dot{y}_0,
\label{eq:defcondinitadim}
\eeq
where the dot means the derivation with respect to the dimensionless time $\widetilde{t}$. These dimensionless initial conditions are 
consistently of order $\epsilon$. The amplitude equations~\eqref{eq:Amplitude} gives
\beq
\widetilde{x}\left(\widetilde{t}\right) = 2 \epsilon a \cos\left(\widetilde{t}+\phi\right),\qquad \widetilde{y}\left(\widetilde{t}\right) = 2 \epsilon b \cos\left(\widetilde{t}+\psi\right),
\label{eq:soluadim}
\eeq
where $a = \sqrt{N(1 - \chi)}$ and $b = \sqrt{N \chi}$. The initial conditions for the dynamical variables $\chi$, $\phi$ and $\psi$ are given by Eqn.~\eqref{eq:defcondinitadim}, and reads
\beq
\begin{cases} \xt = 2 \epsilon \sqrt{N(1 - \chi_0)} \cos\phi_0, \\ \dxt = -2 \epsilon \sqrt{N(1 - \chi_0)} \sin\phi_0, \end{cases} \quad  \begin{cases} \yt = 2 \epsilon \sqrt{N \chi_0} \cos\psi_0, \\ \dyt = -2 \epsilon \sqrt{N \chi_0} \sin\psi_0. \end{cases}
\label{eq:condinitadim}
\eeq
A simple manipulation gives
\beq
\begin{cases} \xt^2 + \dxt^2 = 4 \epsilon^2 N(1 - \chi_0), \\  \yt^2 + \dyt^2 = 4 \epsilon^2 N \chi_0, \end{cases}
\label{eq:manip1}
\eeq
therefore
\beq
 \xt^2 + \dxt^2  + \yt^2 + \dyt^2 = 4 \epsilon^2 N.
\label{eq:manip2}
\eeq
If we define the small parameter $\epsilon$ as 
\beq
 \epsilon \equiv \frac{\sqrt{ \xt^2 + \dxt^2  + \yt^2 + \dyt^2}}{2} \ll 1,
\label{eq:valeps}
\eeq
we can take $N = 1$ in all generality. Physically, the small parameter is the ratio between the initial energy and the depth $U_0$ of the well.

We then get
\beq
 \chi_0 \equiv \frac{\yt^2 + \dyt^2}{ \xt^2 + \dxt^2  + \yt^2 + \dyt^2} , 
\label{eq:valchi}
\eeq
which is consistent with the requirement $0 < \chi_0 <1$. Moreover, since we may write
\beq
 \chi_0 \equiv \frac{\yt^2 + \dyt^2}{ 4 \epsilon ^2}, \qquad   1 - \chi_0 = \frac{\xt^2 + \dxt^2}{4 \epsilon ^2},
\label{eq:valchibis}
\eeq
we see that injecting these expressions in the system~\eqref{eq:condinitadim} we get consistent real values for the phases, since the relevant trigonometric functions range between $-1$ and $1$.

\section{Measuring the amplitude and the phase}
\label{App:Measuring}

The calculation of the variables $\chi$ and $\theta$ from the simulations data follows from the assumption that the rough data $x(t)$ and $y(t)$ may be written
\beq
\begin{cases} x(t) = 2 \epsilon a \cos(t + \phi) = 2 \epsilon a (\cos t \cos \phi - \sin t\sin\phi), \\ y(t) = 2 \epsilon b \cos(t + \psi) = 2 \epsilon b (\cos t \cos \psi - \sin t\sin\psi),\end{cases}
\label{eq:dataana}
\eeq
where $(a,b)$ are slowly varying amplitudes and $(\phi,\psi)$ are slowly varying phases, with characteristic periods much larger than $2 \pi$ which is the quick time period. An example of the signals $x(t)$ and $y(t)$ is displayed in Fig.~\ref{fig:evoltimedata}, for initial conditions that are consistent with a small value of $\epsilon$ (actually, $\epsilon = 0.224$, which is not that small). The time evolution of both signals obviously validates the assumption of a slow variation of the amplitudes. To calculate the amplitudes from the raw simulations data, we extract the local maxima, and then we build from these set an interpolation function in order to get a smooth function for the amplitude.  Doing this, we get the slowly varying function
\beq
\chi(t) = \frac{b^2(t)}{a^2(t) + b^2(t)}.
\label{eq:chinum}
\eeq

To get the slowly varying phases, we multiply the raw numerical data by either $\cos t$ or $\sin t$, and take the average on the fast time variable. For example, we numerically integrate the simulations data to calculate
\beq
\langle X_C \rangle \equiv \frac{1}{2\pi}\int_0^{2\pi} x(t) \cos t \dd t = \epsilon a \cos\phi,
\label{eq:phinum}
\eeq
since the slowly varying functions $a$ and $\phi$ may be considered as constant for the integration. Since we already know the amplitude $a$ as a (slow) function of time, we thus get $\cos \phi$. Replacing $\cos t$ by $\sin t$ in~\eqref{eq:phinum}, we get $\sin \phi$ and doing the same work on $y(t)$ we get $\cos \psi$ and $\sin \psi$ as (slow) functions of time. To get smooth functions, it is convenient to define complex variables $Z_\phi = \cos \phi + i \sin \phi$ and $Z_\psi = \cos \psi + i \sin \psi$, so that the variable $\theta$ is obtained as
\beq
\theta = \mathrm{Arg}\left(\frac{Z_\psi^2}{Z_\phi^2}\right).
\label{eq:thetanum}
\eeq
The numerical algorithms that calculate an angle as the argument of a complex number are the less sensitive to noise, which justifies their use.



\bibliography{Biblio/Biblio_Zigzag,Biblio/Biblio_D1D,Biblio/Biblio_Groupe,Biblio/Biblio_BifNoise,Biblio/Biblio_DiffGen,Biblio/Biblio_Potentiel,Biblio/Biblio_SDHamiltoniens,Footnotes}



\end{document}